\documentclass[usenatbib]{mn2e}
\usepackage{graphicx,times}
\usepackage{bm}
\usepackage{epsf}
\usepackage{amsmath}
\usepackage{amssymb}
\date{\today,~ $ $Revision: 0.9 $ $}
\def\la{\langle}
\def\ra{\rangle}
\def\n{\noindent}
\def\be{\begin{equation}}
\def\ee{\end{equation}}
\def\ben{\begin{eqnarray}}
\def\een{\end{eqnarray}}
\def\nn{\nonumber}
\def\oh{\hat\Omega}
\def\myC{{\cal C}}
\def\myB{{\cal B}}
\def\bk{{\bf k}}

\def\br{{\bf r}}
\def\rH{\rm H}
\def\at{\Theta}

\def\myC{{\cal C}}
\def\myB{{\cal B}}

\def\bk{{\bf k}}

\def\bx{{\bf x}}
\def\2p{{(2\pi)^2}}

\def\be{\begin{equation}}
\def\ee{\end{equation}}
\def\beq{\begin{equation}}
\def\eeq{\end{equation}}
\def\ben{\begin{eqnarray}}
\def\een{\end{eqnarray}}

\def\oh{{\hat\Omega}}
\def\nn{{\nonumber}}

\newcommand{\beqa}{\begin{eqnarray}}
\newcommand{\eeqa}{\end{eqnarray}}

\newcommand{\rSZ}{{\rm SZ}}
\newcommand{\myf}{{\Theta}}

\begin{document}
\onecolumn
\title[The Morphology of the $y$-Sky]
{The Morphology of the Thermal Sunyaev-Zel'dovich Sky}
\author[Munshi et al.]
{Dipak Munshi$^{1}$, Joseph Smidt$^{2}$, Shahab Joudaki$^{2}$, Peter Coles$^{1}$ \\
$^{1}$School of Physics and Astronomy, Cardiff University, Queen's
Buildings, 5 The Parade, Cardiff, CF24 3AA, UK\\
$^{2}$ Department of Physics and Astronomy, University of California, Irvine, CA 92697}
\maketitle
\begin{abstract}
At high angular frequencies, beyond the damping tail of the primary
CMB power spectrum, the thermal Sunyaev-Zel'dovich (tSZ) effect
constitutes the dominant signal in the CMB sky. The tSZ effect is
caused by large scale pressure fluctuations in the baryonic
distribution in the Universe so its statistical properties can
provide estimates of corresponding properties of the projected 3D
pressure fluctuations. The power spectrum of the tSZ is a sensitive
probe of the amplitude of density fluctuations, and the bispectrum
can be used to separate the bias associated with the pressure. The
bispectrum is typically probed with its one-point real-space
analogue, the skewness. In addition to the ordinary skewness the
morphological properties, as probed by the well known Minkowski
Functionals (MFs), also require the generalized one-point
statistics, which at the lowest order are identical to the
generalized skewness parameters. The concept of generalized skewness
parameters can be further extended to define a set of three
associated generalized {\it skew-spectra}. We use these skew-spectra
to probe the morphology of the tSZ sky or the $y$-sky. We show how
these power spectra can be recovered from the data in the presence
of an arbitrary mask and noise templates using the well known
Pseudo-$C_l$ (PCL) approach for arbitrary beam shape. We also employ
an approach based on the halo model to compute the tSZ bispectrum.
The bispectrum from each of these models is then used to construct
the generalized skew-spectra. We consider the performance of 
an all-sky survey with Planck-type noise and compare the results
against a noise-free ideal experiment using a range of smoothing
angles. We find that the skew-spectra can be estimated with very
high signal-to-noise ratio from future frequency cleaned tSZ maps
that will be available from experiments such as Planck. This will
allow their {\it mode by mode} estimation for a wide range of
angular frequencies $l$ and will help us to differentiate them from
various other sources of non-Gaussianity.
\end{abstract}
\section{Introduction}
Measurements of cosmological parameters from Cosmic Microwave
Background (CMB) surveys such as those performed by satellite
missions like  WMAP\footnote{http://wmap.gsfc.nasa.gov/} and
Planck\footnote{http://www.rssd.esa.int/Planck} can not only provide
a very accurate picture of the background geometry and dynamics of
the Universe but can also reveal much more detailed information
about ongoing physical processes. Indeed, with all-sky coverage and
wide range of previously uncharted frequencies the data provided by
experiments such as Planck will produce ``secondary'' science which
is arguably as valuable as the ``primary'' science; see e.g.
\cite{AMS08} for a recent review. In any case accurate
modeling of secondary non-Gaussianities is required
to avoid $20\%-30\%$ constraint degradations in future CMB datasets such
as Planck and CMBPol\footnote{http://cmbpol.uchicago.edu/}\citep{Smidt10}.

The large-scale properties of hot intergalactic gas can be probed
through the multifrequency observations. Inverse-Compton scattering
of CMB photons known as thermal Sunyaev-Zel'dovich effect (tSZ)
 leaves a characteristic distortion
pattern in the CMB spectrum \cite{SZ72,SZ80,Bir99,Rep95}. The
fluctuation of this distortion across the sky as probed by CMB
observations can provide valuable clues to the fluctuations of the
gas density and temperature. In the (low frequency) Rayleigh-Jeans
(RJ) regime it produces constant decrement, while there is an
increment at high frequencies; in between there is a  a null (around
$217$GHz). This characteristic behavior is a potential tool for the
separation of tSZ from the other contribitions to the temperature
anisotropy. Based on accurate knowledge of the tSZ and CMB spectrum,
foreground removal techniques have been developed to isolate the tSZ
signal in the presence of primary anisotropy and instrumental noise.
These techniques are extremely effective in the subtraction of
primary anisotropies due to its well understood (perfect black body)
frequency dependence and almost exactly Gaussian statistical
behavior \citep{Leach08,BG99,DCP03}.
The tSZ effect is now routinely imaged in massive individual galaxy
clusters where the temperature of the scattering medium can reach
$10{\rm keV}$. This in effect produces a change in CMB temperature
of order $1$mK at RJ  wavelengths. Individual galaxy cluster tSZ
images have a variety of astrophysical and cosmological applications
including direct measurement of the angular diameter distance to the
cluster through a combined analysis of X-ray data and measurement of
the gas mass which can be useful in estimation of baryon fraction of
the Universe. The High Frequency Instrument (HFI) of Planck, in
particular, has been designed with bands centered at the minimum,
the null, and the maximum of the thermal SZ (tSZ) emission. The
extraction of frequency cleaned CMB and tSZ maps of a catalog of
galaxy clusters selected by their tSZ effect are part of the
scientific program of Planck. Here we are however also interested in
the general intergalactic medium (IGM) where the gas is expected to
be at $\le {\rm 1keV}$ in mild overdensites which leads to CMB
contributions in the $\mu$K range. In this work we primarily focus
on the statistical study of wide-field CMB data where tSZ effects
lead to anisotropies in the temperature distribution due to both
resolved and unresolved galaxy clusters, keeping in mind that the
thermal tSZ contribution is the dominant signal beyond the damping
tail of the primary anisotropy power spectrum. We primarily focus on
analytical modelling morphological properties of the tSZ
fluctuations using Minkowski Functionals (MFs).

Modelling of the lower order statistics of the tSZ effect can be
performed using either analytical or numerical approaches. Authors
using analytical approaches  \citep{Sj00,Zp01,KS02,ZS07,AC1,AC2}
have generally used the halo model \citep{CooSeth02}.  To trace the
tSZ effect due to photo-ionized gas outside collapsed halos, see
e.g. \cite{AC2}. A second-order perturbative formalism is used to
model the bispectrum in this approach. The gas will typically be at
a temperature similar to the ionization energy of hydrogen and
helium. The bias associated with the pressure fluctuations is
assumed to be redshift dependent. For tSZ effect from material
within collapsed halos the shock-heated gas is typically assumed to
be in hydrodynamic equilibrium in virialized halos. The statistical
description of halos that include the number count distribution are
assumed to be described by Press-Schechter formalism \citep{PS74}.
The radial profile of such halos are assumed to be that of NFW
\citep{NFW96}. These ingredients are sufficient for modeling of tSZ
effect from collapsed halos \citep{AC1}.

In addition to analytical modeling the numerical simulation of tSZ
plays an important role in our understanding of the physics involved
\citep{Persi95,Ref00,Sel01,Spr01,White02,Lin04, Z04,
Cao07,Ron07,Hal09,Hal07,dasilva99}. Some of these studies
incorporate complications from additional gas physics such as
radiative cooling, preheating and SN/AGN feedback, at least to a
certain extent. The inputs are otherwise difficult to incorporate in
any analytical calculations. On the other hand the simulations are
limited in their dynamic range and can benefit from analytical
insights.

The tSZ power spectrum is known to be a sensitive probe of the
amplitude of density fluctuations. Higher order statistics, such as
the skewness or the bispectrum we study here, can provide
independent estimates of the bias associated with the baryonic
pressure. Typically a collapsed three-point statistics such as the
(one-point) skewness is employed for this purpose. The skewness
compresses all available information in the bispectrum. The recently
proposed skew-spectrum is a power spectrum associated with the
bispectrum \citep{MuHe10} which is useful to probe non-Gaussianity
as a function of scale or the harmonics $l$. In addition to the
lower-order multispectra, morphological statistics such as the
Minkowski Functionals (MFs) carry independent information of
non-Gaussianity and have been studied extensively in the literature
\citep{Hk08,HKM06,HTS03,Hk02,Hk03}. At the lowest order the MFs are
completely described by a set of three different skewness parameters
that describe complete set of three Minkowski functionals in 2D. We
extended these generalized skewness parameters to their
corresponding skew-spectra. These skew-spectra sample the bispectra
with varying weights and carry independent information. The
estimation of these skew-spectra can be done at relatively modest
computational cost. The skew-spectra can be constructed by
cross-correlating suitable maps that are constructed from the
original (frequency cleaned) tSZ maps. The construction involves
differential operations on beam smoothed pixelised maps that can
also be performed in the harmonic domain. The estimators for
skew-spectra can be constructed following the general principle of
power spectrum estimation. We use the well known Pseudo $C_l$s (PCL)
approach developed by \citep{Hiv}. It can handle arbitrary sky
coverage and arbitrary noise characteristics and beam patterns. We
use the PCL approach to construct the variance in the estimators as
well as to compute their cross-correlations. These results are
accurate for surveys with all-sky coverage and can also be modified
to take into account partial sky coverage, using the flat sky
approximation.

To relate to the experimental scenarios we will consider an experimental setup similar to the ongoing all-sky experiment Planck
\citep{PC06}. We consider the range of harmonics $(2,2000)$. However, the results presented here are also applicable to
smaller surveys such as Arcminute Cosmology Bolometer Array
Receiver (ACBAR; \cite{Run03})\footnote{http://cosmology.berkeley.edu/group/swlh/acbar/}
which covers the $\ell$-range $(2000,3000)$. ACBAR is a
multifrequency millimeter-wave receiver designed for observations of
the cosmic microwave background and the Sunyaev-Zel'dovich effect.
The ACBAR focal plane consists of a 16 pixel, background-limited,
240 mK bolometer array that can be configured to observe
simultaneously at $150$, $220$, $280$, and $350$ GHz with $4'$-$5'$
FWHM. Together with Planck these two experiments will cover the entire range of
$\ell$ values up to $3000$. In addition to Planck we also consider a noise free ideal set up
for the  range of $\ell$ values $(2,2000)$ as a reference.

The paper is organized as follows. In \textsection\ref{sec:low} we
review the details of the analytical models involving the power
spectrum and the bispectrum of the tSZ effect. In
\textsection\ref{sec:MF_intro} we present the skewness parameters
that can be used to study the Minkowski Functionals. In
\textsection\ref{sec:MF} we define the triplets of Generalized
Minkowski Functionals. We propose the use of skew-spectra associated
with individual generalised skewness parameters that carry more
information than the ordinary MFs. We show how these skew-spectra
are associated with the MFs and are related to the generalised
skewness parameters. Next in \textsection\ref{sec:scatter} we
present the estimators that can be used to extract the skew-spectra
from a noisy data set in the presence of observational mask and
arbitrary beam. We also obtain the signal-to-noise for these
skew-spectra for realistic observational scenarios. Finally,
\textsection\ref{sec:conclu} is devoted to a discussion of our
results and future prospects.

The particular cosmology that we will adopt for numerical
study is specified by the following parameter values (to be introduced later):
$\Omega_\Lambda = 0.741, h=0.72, \Omega_b = 0.044, \Omega_{\rm CDM} = 0.215,
\Omega_{\rm M} = \Omega_b+\Omega_{\rm CDM}, n_s = 0.964, w_0 = -1, w_a = 0,
\sigma_8 = 0.803, \Omega_\nu = 0$.
\section{Lower order Statistics of the tSZ effect}
\label{sec:low}
In this section we briefly review the two different approaches that are
commonly used to model the tSZ effect (for more details see \citep{CBS05}
and the references there in).  We will use these models
later to study the morphological properties of the SZ effect. The
tSZ temperature fluctuation $\Theta^{\rm SZ}(\oh,\nu)= \delta
T(\oh,\nu) /T_{\rm CMB}$ is given by the (opacity-weighted) pressure
fluctuations i.e. \be \Theta^{\rm SZ}(\oh,\nu) = {\delta T(\oh)/
T_{\rm CMB}} \equiv g_{\nu}(x)y(\oh) = g_\nu(x)\int_0^{r_0}\;d\eta\;
a(\eta)\;{\sigma_T / m_e}\;\pi_e(\oh,\eta);\ee we will use the symbol
$\bx$ to represent a comoving coordinate. The electron pressure is
denoted as $\pi_e = n_ek_BT_e$. Here $y(\oh)$ is the Compton
$y$-parameter, $\sigma_T$ is the Thomson cross-section, $k_B$ is the
Boltzman constant, $m_e$ is the electron rest mass, $n_e$ is the
electron number density and $T_e$ the electron temperature. The
function $g_\nu(x)$ encodes the frequency dependence of the tSZ
anisotropies. It relates the temperature fluctuations at a frequency
$\nu$ with the Compton parameter $y$, i.e. $g_\nu(x) = x\coth
\left({x/2}\right) -4$ and $x= {(h\nu / k_B T_{\rm CMB})} = {\nu /
(56.84 {\rm GHz})}$. In the low frequency part of the spectrum
$g_\nu(x) = -2$, for $x \ll 1$, here x is the dimensionless
frequency as defined above. Unless stated otherwise all our results
will be for the low frequency limiting case. The conformal time
$\eta$ can be expressed in terms of the cosmological density
parameters $\Omega_{\rm M}$ $\Omega_{\rm K}$ and $\Omega_{\Lambda}$
by the following expression: $\eta(z) = \int_0^z {dz'/ {\rm H}(z')}$
and ${\rm E}^2(z) = {\rH^2(z) / \rH_0^2} = [\Omega_{\rm M}(1+z)^3 +
\Omega_{\rm K}(1+z)^2 + \Omega_{\Lambda}]$. Introducing the $\tau$
the Thompson optical depth and the fractional fluctuations in
pressure as $\pi = \delta p_e / \rho_e$ we can write
$\Theta^{\rm SZ}(\oh,\nu) = \int_0^{r_0}  dr \dot \tau \pi(\oh,r)$. Here
$r$ is the comving distance, $\tau$ is the Thomson optical depth,
and overdots represent derivatives with respect to $r$.

We will primarily be working in the Fourier domain. The
three-dimensional electron pressure $\pi$ can be decomposed into its
Fourier coefficients using the following convention that we will
follow: $\delta\pi(\bx) = (\pi(\bx)-\la\pi(\bx)\ra)/ \la\pi(\bx)\ra$ with ${\delta\pi}(\bk) =
\int d^3 \bx \;{\delta\pi}(\bx) \exp[{-i\bk\cdot\bx}]$. The projected
statistics that we will consider can be related to the 3D statistics
which are defined by the following expressions which specifies the
power spectrum $P_{\pi}(k)$, the bispectrum $B_{\pi}(k_1,k_2,k_3)$ and the
trispectrum $T_{\pi}(k_1,k_2,k_3,k_4)$ in terms of the Fourier
coefficients:
\ben
&& \la \delta\pi(\bk_1,r_1)\delta\pi(\bk_2,r_2) \ra_c = (2\pi)^3 P_{\pi}(k_1;r_1,r_2) \delta_{3D}(\bk_1+\bk_2) \label{eq:ps} \\
&& \la \delta\pi(\bk_1,r_1)\delta\pi(\bk_2,r_2)\delta\pi(\bk_3,r_3) \ra_c = (2\pi)^3 B_{\pi}(k_1,k_2,k_3;r_i)\delta_{3D}(\bk_1+\bk_2+\bk_3) \label{eq:bs} \\
&& \la \delta\pi(\bk_1,r_1)\delta\pi(\bk_2,r_2)\delta\pi(\bk_3,r_3)\delta\pi(\bk_4,r_4) \ra_c = (2\pi)^3 T_{\pi}(k_1,k_2,k_3,k_4;r_i)\delta_{3D}(\bk_1+\bk_2+\bk_3+\bk_4)
\label{eq:ts}
\een
\n Our aim is to relate the statistics of pressure fluctuations
$\delta\pi(\bx)$ to those of the underlying density fluctuations $\delta(\bx)$.
The hierarchy of higher-order correlation functions can be defined
using a similar set of equations. The corresponding multispectra
will be denoted with a subscript $\delta$ i.e. $B_{\delta}$ and
$T_{\delta}$ will represent the bi- and trispectrum of the
underlying density contrast and the power spectrum will be denoted
as $P_{\delta}$. The $\delta_{3D}$ functions represent the 3D Dirac
delta function and represent translational invariance of
corresponding correlation hierarchy in real space. The unequal time
correlators appearing in the expressions of the multispectra  will
typically collapse to equal time correlators due to the use of
Limber approximations that we will discuss next.
\begin{figure}
\begin{center}
{\epsfxsize=6.75 cm \epsfysize=6 cm {\epsfbox[276 450 587 709]{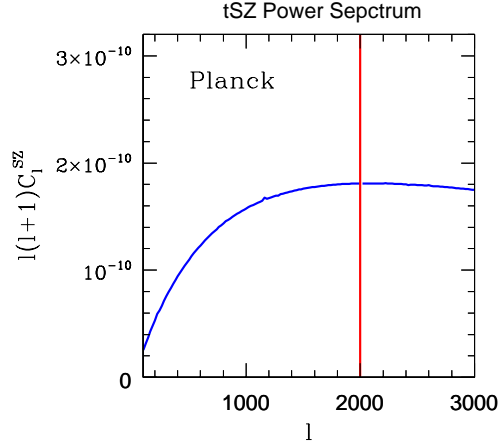}}}
\end{center}
\caption{The power spectrum of SZ is depicted for our choice of cosmological parameters. The halo model
was used for the computation
of the power spectrum. See section \textsection\ref{sec:halo} for a short discussion and relevant parameter values.
For an all-sky Planck type experiment we consider the range $l=(2,2000)$ }
\label{fig:ps}
\end{figure}
\subsection{Linking projected statistics with 3D statistics}
\label{subsec:2D3D}
The spherical harmonics decomposition of the projected field
$\Theta^{\rm SZ}(\oh)$ will be represented as $\at^{\rSZ}_{lm}$ can
be expressed in terms of a weight  $W^{SZ}(r)$. We will henceforth
suppress the frequency $\nu$ dependence of $a^{\rSZ}_{lm}$ as well
as $\Theta^{\rSZ}(\oh)$. The line-of-sight integration over $r$
projects the 3D  $\delta_{lm}(\bk,r)$ onto the harmonics
$a^{\rSZ}_{lm}$:
\be
\at^{\rSZ}_{lm} =  \int d\oh \; Y^*_{lm}(\oh) \; \Theta^{\rSZ}(\oh)
= \sqrt{2 \over \pi} \int_0^{r_0} W^{\rm SZ}(r) \int k^2dk \sum_{lm} j_l(kr)Y^*_{lm}(\oh) \delta_{lm}(\bk,r); \quad  W^{SZ}(r)= -2b_{\pi}(r)\dot \tau; \quad \dot \tau = c \sigma_T n_e(z)
\label{eq:harm}
\ee
\n Here $Y_{lm}(\oh)$ represents a spherical harmonic and $j_l(kr)$
is the spherical Bessel function of order $l$. The weight
$W^{SZ}(r)$ is introduced by projecting the 3D statistics onto the
sky. We use the following relation to relate $\delta_{lm}(\bk,r)$ in
the spherical basis with $\delta(\bk,r)$ in Fourier basis to obtain
that \be \delta_{lm}(\bk) = {(2\pi)^{-3/2}} \int d\oh\;
Y_{lm}(\oh)\; \delta(\bk)\ee. We can use this expression next in
Eq.(\ref{eq:harm}) and use the definition of power spectrum from
Eq.(\ref{eq:ps}). Using the integral representation of the Dirac's
delta function $\delta_{3D}({\bk}) = {(2\pi)^{-3}} \int
\exp(i\bk\cdot\bx)d^3\bx$; in association with the Rayleigh's
expansion of the plane wave in terms of spherical harmonics
$\exp(i\bk\cdot\bx) = 4\pi\sum_{lm} i^l
j_l(kx)Y_{lm}(\oh_k)Y_{lm}(\oh)$. Using these expression we arrive
at the following expression which projects the 3D power spectrum
$P(k,r_1,r_2)$ to the spherical sky \citep{AC2}:
\be
\myC_l^{\rSZ} = {2 \over \pi} \int W^{\rm SZ}(r_1) dr_1 \int W^{\rm SZ}(r_2)dr_2 \int dk k^2 j_{l_1}(kr_1) j_{l_2}(kr_2) P_{\pi}(k,r_1,r_2)
\ee
\n The following form of Limber's approximation
\citep{Limb54,LoAf08} is remarkably accurate at large angular
scales:  $l \le 100$: $\int F(k) k^2 dk j_{l_1}(k_1r)k_{l_2}(k_2r) =
{(\pi/2r^2)} F\left ({l/r} \right )\delta_D(r_1-r_2)$. The use of
this form of Limber's approximation greatly simplifies the
subsequent results by reducing the correlator of multiple time
slices to single time correlators  \citep{AC2}.
\be
\myC_l^{\rSZ} =\la \at_{lm}^{\rm SZ}\at_{lm}^{\rm SZ*} \ra =  {} \int_0^{r_0}\; dr\;{[W^{\rSZ}(r)]^2 \over d_A^2(r) } P_{\pi}\left ({l \over d_A(r)}; r \right )
\label{eq:pro_ps}
\ee
\begin{figure}
\begin{center}
{\epsfxsize=6 cm \epsfysize=6 cm {\epsfbox[27 426 315 709]{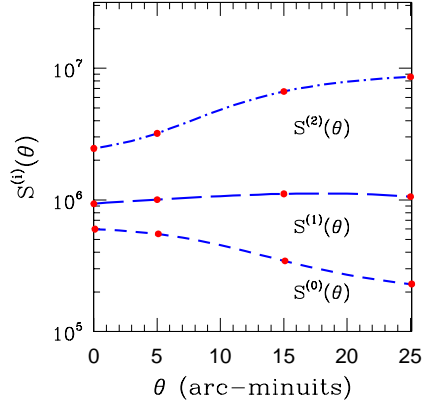}}}
\end{center}
\caption{The three one-point skewness parameters $S^{(0)}$ (short-dashed), $~S^{(1)}$ (long-dashed) and
$S^{(2)}$ (dot-dashed) are plotted as a functions of
the FWHM $\theta$ for a Gaussian beam. The results plotted are for ideal surveys without instrumental noise.
The skewness parameter $S^{(0)}$
is the ordinary skewness describing the departure from non-Gaussianity at the lowest order. The points are results from
numerical computations using the halo model and the lines denote smooth fits to the results. The one point skewness parameters
$S^{(i)}$ is related to the skew-spectra $S_l^{(i)}$ through the following expression: $S^{(i)} =\sum_l (2l+1)S_l^{(i)}$. }
\label{fig:one}
\end{figure}
\n
Here $d_A(r)$ is the angular diameter distance in terms of the comoving radial distance $r$.
The Power spectrum $\myC_l^{\rSZ}$ is plotted in Fig-\ref{fig:ps} as a function of the harmonics $l$.

The bispectrum in the harmonic domain is represented by the following three-point correlation
function:
\be
B^{\rm SZ}_{l_1l_2l_3}\equiv \sum_{mm'm''}
\left ( \begin{array}{ c c c }
     l & l' & l'' \\
     m & m' & m''
  \end{array} \right )\la \at^{\rm SZ}_{lm}\at^{\rm SZ}_{l'm'}\at^{\rm SZ}_{l''m''}\ra_c
\label{eq:def_3j}
\ee
\n The matrix above in Eq.(\ref{eq:def_3j}) represents the Wigner
$3j$ symbols \citep{Ed68} and $B^{\rm SZ}_{l_1l_2l_3}$ defines the
tSZ bispectrum in the spherical sky. The equation linking the
projected bispectrum $B^{\rm SZ}_{l_1l_2l_3}$ and its 3D analogue
$B_{\pi}(k_1,k_2,k_3)$, defined in Eq.(\ref{eq:bs}), can be derived
following exactly the same procedure. We quote here the final
expression which takes the following form \citep{AC2}:
\be
B^{\rm SZ}_{l_1l_2l_3} = I_{l_1l_2l_3}\int dr \;{[W^{\rSZ}(r)]^3 \over d_A^3(r) }
B_{\pi} \left ( {l_1\over d_A(r)}{l_2\over d_A(r)}{l_3\over d_A(r)};r \right ); \quad
I_{l_1l_2l_3} =  \sqrt{(2l_1+1)(2l_2+1)(2l_3+1) \over 4 \pi}
\label{eq:pro_bs}
\ee
Next we describe models for the 3D power spectrum $P_{\pi}(k;r)$ and
bispectrum $B_{\pi}(k,k',k'';r)$ that will be used for computation
of the projected SZ versions $\myC_l^{\rSZ}$ and $B^{\rm
SZ}_{l_1l_2l_3}$. Modelling of the tSZ statistics  requires a model
for the clustering of underlying dark matter distributions $\delta$.
We will employ extensions of perturbative calculations as well as
halo model predictions to model the underlying statistics of dark
matter clustering. We will describe these model in following
subsections. We have shown the angular power spectrum $\myC_l^{SZ}$
for SZ effect in Figure-\ref{fig:ps} as a function of harmonic
number $l$. For the computation of this power spectra a halo model
approach was used which will be discussed in
\textsection\ref{sec:halo}.
As we discussed in the Introduction, modelling the tSZ statistics
involves modeling the underlying dark matter clustering and its
relationship with the baryonic clustering and thence the fluctuation
in baryonic pressure. This modeling has so far been performed using
two complimentary approach. In the simpler approach the clustering
of dark matter is described using second order perturbation theory
or its extensions. The clustering of baryons are described using a
biasing scheme. These inputs are sufficiently accurate especially at
larger length scales to model the SZ sky.  We will consider the
linear biasing model first.
\begin{figure}
\begin{center}
{\epsfxsize=15 cm \epsfysize=5 cm {\epsfbox[49 537 577 709]{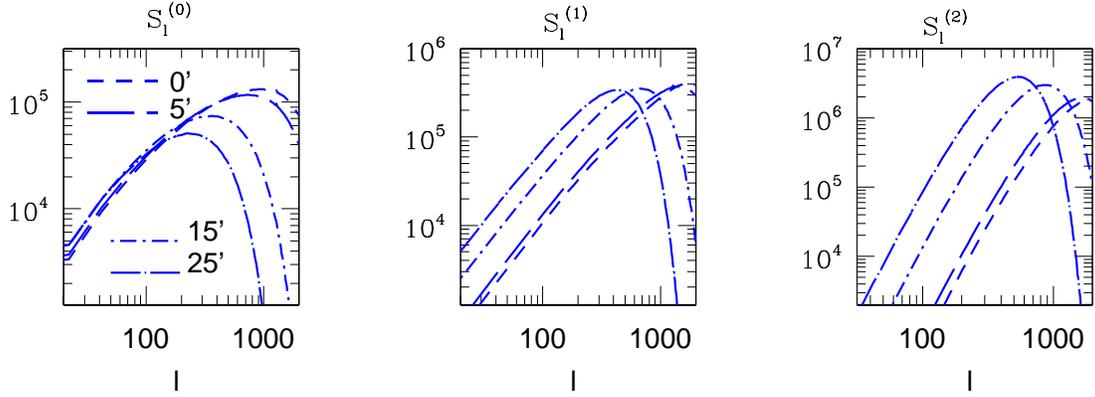}}}
\end{center}
\caption{The absolute values of skew-spectra are plotted as a function of the harmonic $l$. From left to right
the plots correspond to $S_l^{(0)}$, $S_l^{(0)}$ and $S_l^{(0)}$ respectively. We consider
 four different beams with FWHM = $0'$, $5'$, $15'$ and $25'$ respectively as indicated.
These results correspond to {\em ideal noise-free} experiments and the resolution is taken to be $l_{max}=2000$. The results
are obtained using halo model prescription.}
\label{fig:skew_spec2a}
\end{figure}
\begin{figure}
\begin{center}
{\epsfxsize=15 cm \epsfysize=5 cm {\epsfbox[27 517 587 709]{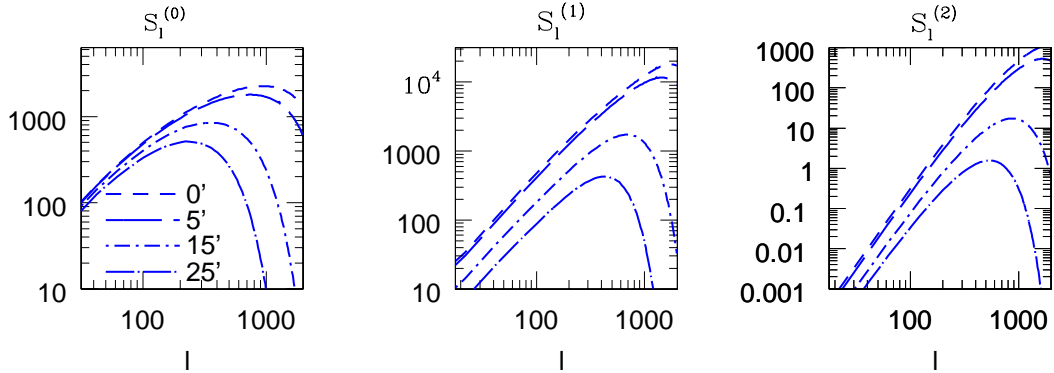}}}
\end{center}
\caption{The absolute values of skew-spectra are plotted as a function of the harmonic $l$. From left to right
the plots correspond to $S^{(0)}$, $S^{(1)}$ and $S^{(2)}$ respectively. We consider
 four different beams with FWHM = $0'$, $5'$, $15'$ and $25'$ respectively as indicated.
These results include Planck-type noise and correspond to an all-sky survey.}
\label{fig:skew_spec2b}
\end{figure}
\subsection{Linear Biasing and Perturbative Approach}
\label{subsec:lin_bias}
In a perturbative approach the gravitational dynamics is analyzed by expanding the large scale density
field $\delta$ in a perturbative series and a redshift dependent linear biasing is assumed\citep{GS99a,GS99b}. 
It is valid as long as the variance of the density contrast is smaller than unity and such a treatment is
suitable for diffuse tSZ component \citep{Han05}. For large smoothing scales such calculations
 can provide valuable insight to gravitational dynamics. The perturbative bispectrum $B_{\delta}(k_1,k_2,k_3;r)$ 
for the density contracst $\delta$ from such calculations is given by \citep{Peebles71}:
\ben
&& B_{\delta}(k_1,k_2,k_3;r) = F_2(k_1,k_2)P_{\delta}(k_1,r)P_{\delta}(k_2,r) + {\rm cyc.perm.} \\
&& F_2(k_1,k_2) = {1-{2 \over 7}\Omega_M^{-2/63}} + {{\bf k}_1\cdot {\bf k}_2 \over 2 k_1 k_2 }\left ( {k_1 \over k_2} + {k_2\over k_1}\right ) + {2 \over 7} \Omega_M^{-2/63}{({\bf k}_1\cdot{\bf k}_2)^2 \over k_1^2k_2^2}
\label{eq:pert}
\een
In the deeply nonlinear regime we do not have a complete analytical
model of gravitational clustering. Typically, plausible
approximations that include the halo model (to be discussed next)
are used for this purpose, with varying degree of success. One such
ansatz is the well known {\em hierarchical ansatz} that assumes that
the higher order correlation functions can be constructed from the
products of two point correlation functions \citep{Bernardreview02}.
The different tree diagrams thus constructed are all represented by
various {\em hierarchical  amplitudes} which are left arbitrary. The
bispectrum  in such a hierarchical scenario takes the following
form: 
\be
B_{\delta}(k_1,k_2,k_3;r) = Q_3(n) \left [ P(k_1)P(k_2) + {\rm
cyc.perm.} \right ]; \quad Q_3(n) = {[4 - 2^n] / [1 + 2^{n+1}] }.
\ee
The expression for $Q_3(n)$ is adopted from \citep{SF99}, $n$ here is the
local linear power spectral index. For a
generic power spectrum one can replace $n$ with the  local linear
power spectral index at $(k_1+k_2+k_3)/3$ \citep{Hui99}. More
elaborate schemes that interpolate the quasilinear regime and the
highly non-linear regime have been devised in recent years
\cite{SC01}. Though we will restrict ourselves to the computation of
the bispectrum in this paper, the approach can also be extended
beyond third order. The hierarchical ansatz captures the salient
features of non-linear clustering; the expressions are relatively
simpler than the full perturbative expressions and have been tested
in variety of cosmological contexts. In connecting the statistics of
the electron pressure field $\pi$ namely the power spectrum
$P_{\pi}(k;r)$ and $B_{\pi}(k,k',k'';r)$ we will use the following
local biasing model $P_{\pi} (k,r) = b_{\pi}(r)^2 P_{\delta}(k,r)$
and $B_{\pi}(k,k',k'') = b_{\pi}(r)^3  B_{\delta}(k,k,k;r)$.
Following \citep{GS99a,GS99b,AC1,AC2} the bias is chosen to be of
the following form $b_{\pi}(r) = {b_{\pi}(0)/(1+z)}$ with
$b_{\pi}(0)=   {k_B T_e(0)/(m_ec^2 b_{\delta})}$. In our calculation we 
take $b_{\pi}(r) = {0.0039/(1+z)}$. These 3D Fourier
statistics will be the ingredients for computation of projected
statistics on the surface of the sky using Eq.(\ref{eq:pro_ps}) and
Eq.(\ref{eq:pro_bs}).
\subsection{Halo Model and tSZ: A brief Review}
\label{sec:halo}
The halo model description of large scale structure  relies on
modeling of the clustering of halos and predictions from
perturbative calculation to model the non-linear correlation
functions. In the context of halo model the dark matter is assumed
to be in collapsed halos which are characterized by their mass, A brief Review
radial profile, halo occupation numbers and underlying correlation
hierarchy. The baryonic fluid is assumed to be in equilibrium with
the dark matter profile and is characterized by a specific equation
of state. The use of halo model is well established in cosmology and
has a long history \citep{CooSeth02}. It has been used relatively
recently is modeling of clustering of galaxies, weak lensing
observables and statistics of CMB secondaries. In the context of tSZ
it has  been used to compute the leading order non-Gaussianity such
as bispectrum as well as higher order statistics such as trispectrum
which is important is computing the covariance of ordinary power
spectrum.
It is known that the halo overdensity at a given position ${\bf x}$,
$\delta^h({\bf x}, M; z)$ can be related to the underlying density
contrast $\delta({\bf x}, z)$ by a Taylor expansion as was shown by
\citep{MoWh96}: 
\be
\delta^h({\bf x}, M; z) =[b_1(M;z) \delta({\bf
x},z) + {1 \over 2} b_2(M,z) \delta^2({\bf x},z) + \dots].
\ee
The expansion coefficients are functions of the threshold $\nu(M;z) =
{\delta_c(z) / \sigma(M,z)}$. Where $\delta_c$ is the threshold for
a spherical over-density to collapse and $\sigma(M,z)$ is r.m.s
fluctuation within a top-hat filter.The parameter $\delta_c(z)$ is
the value of the spherical over-density at which it collapses at a
redshift $z$ and is given by the following functional fit for a
given $\Omega_{\Lambda}$ and $\Omega_M$:
 \be
\delta_c(z) = {3
(12\pi)^{2/3} \over 20} \left [ 1 - {5 \over 936} \ln \left ( 1 +
{{1-\Omega_M \over \Omega_M (1+z)^3}} \right )\right ]
\ee
The rms
fluctuation in the sphere of radius $R$ is constructed from linearly
extrapolated power spectrum $P(k)$ $\sigma^2(M;z) = D^2(z)\int {d
\ln k} {\Delta^2(k)} |W(kR)|^2$ where $W(x)$ is the to-phat window
and $\Delta^2(k)$ is the dimensionless power spectrum is given by
$\Delta^2(k)=k^3P(k)/2\pi^2$. The linear order bias $b_1(M;z)$
introduced  in in this expansion of $\delta_h$ in terms of
underlying density contrast $\delta$ depends on the threshold
$\nu(M;z)$ \citep{MJW97}:
\be
b_1(M;z)= 1 + {(a \nu^2(M;z) - 1)\over \delta_c(z)} + {2p \over 
\delta_c(z) \left \{  1 + [a\nu^2(M;z)]^p \right \}  }.
\ee
The linear
growth factor $D(z)$ is defined by the following expression: $D(z)
\propto {H(z)}{\int_z^{\infty} dz' (1+z') [H(z')]^{-3}}$ and is
normalized to unity at $z=0$ i.e $D(0)=1$. The main two ingredients
in the halo model are the radial profile of the halos and the number
density of halos. The number density is given by : $f(\nu) =
\sqrt{2A^2a^2 / \pi}[1 + (a\nu^2)^{-p}] \exp \left ( -{a\nu^2/
2}\right )$ and the threshold $\nu$. The parameters $(p,a)$ takes the
value $(0,1)$ for Press-Sechter mass function \citep{PS74}. The constant $A$
is determined by imposing the normalization $\int f(\nu)d\nu=1$.
The halo model incorporates perturbative aspects of gravitational
dynamics through the halo-halo correlation hierarchy. The nonlinear
features take direct contribution from the halo profile; the
number-density of haloes also needs to be determined. The total
power spectrum $P^t(k)$ at non-linear scale can be written as
\citep{Sj00} sum of two separate contributions to the power spectrum
$P^t_{\pi}(k,z) = P^{1h}_{\pi}(k,z) + P^{2h}_{\pi}(k,z)$. These represent
contributions from clustering of dark matter particles in two
different halos, i.e. the halo-halo term and a contribution from a
single halo or one-halo  term $P^{1h}$. These terms are in turn
expressed in terms of two integrals:
\be
P^{1h}_{\pi}(k,z)=I^{0}_{2,\pi}(k,z); \quad P^{2h}_{\pi}(k,z)= [I^{(0)}_{1,\pi}(k,z)]^2P_{\delta}(k,z)
\ee
The integrals themselves depend on the number
counts of halos as a function of the concentration parameter $c$ and
the mass $M$.
\begin{eqnarray}
I^{(0)}_{2,\pi}(k;z) = \int dM \int dc {d^2n \over dMdc} |\pi_e(k,| M,c;z)|^2; \quad
I^{(1)}_{1,\pi}(k;z) =   \int dM \int dc {d^2n \over dMdc} b_1(M,z)|\pi_e(k,| M,c;z)|
\end{eqnarray}
The bispectrum can be expressed in an analogus manner:
\begin{eqnarray}
&&B_{\pi}^t(k_1,k_2,k_3)= B_{\pi}^{3h}(k_1,k_2,k_3)+B_{\pi}^{2h}(k_1,k_2,k_3)+B_{\pi}^{1h}(k_1,k_2,k_3)\\
&& B_{\pi}^{1h} = I_3^0(k_1,k_2,k_3); \quad B_{\pi}^{2h}(k_1,k_2,k_3)= I_2^1(k_1,k_2)I_1^0(k_3)P_{\pi}(k_3)+cyc.perm. \\
&& B_{\pi}^{3h} = [2J(k_1,k_2,k_3)I_1^1(k_3)+I^2_1(k_3)]I_1(k_1)I_1(k_2)P_{\delta}(k_1)P_{\pi}(k_2)+ cyc.perm.
\end{eqnarray}
The kernel $J(k_1,k_2,k_3)$ is same as $F_2(k_1,k_2,k_3)$ introduced in Eq.(\ref{eq:pert}). We have introduced the following 
notation above:
\be
I^{(p)}_{q,\pi}(k;z) =   \int dM \int dc {d^2n \over dMdc} b_p(M,z)|\pi_e(k,| M,c;z)|^q.
\ee
The Fourier transform of the halo elctron pressure profile $\pi(r)$ is denoted by $\pi_e(k)$ above:
\be
\pi_e(k) = \int d^3\br \;\pi_e(r)\; \exp(-i\bk\cdot \br) = \int_0^{\infty} 4\pi r^2\;dr\; \pi_e(r) {\sin(kr) \over kr}.
\ee
\n For the dark matter profile we assume $\rho_d(r) \equiv \rho_s
u^{-1}(1+u)^{-2}$ with $u \equiv {r/r_s}$. Here the dark matter is
assumed to be a NFW profile and $r_s$ is a characteristic scale
radius. In the context of the spherical collapse model, the outer
extent of a cluster is taken to be its virial radius $r_v = \left [
{3M / 4 \pi \rho_M(z) \Delta(z) }\right ]$ where $\rho_M(z)$ is the
mean background matter density of the Universe at redshift $z$ and
$\Delta_M(z)$ is the over-density of the halo relative to the
background density. The ratio of the virial radius to the scale
radius is called the concentration parameter $c \equiv r_v/r_s$.
Together $c$ and $M$ determines the dark matter distribution of a
given halo. The gas density profile in terms of the dimensionless
parameter $u$ is $\rho_g(u) \equiv \rho_0 \rho_g(u)$. In this
particular case a polytropic equation of state is typically used.
According to the numerical simulation clusters of a given mass have
a range of concentration parameters. The number density of clusters
with a given mass $M$ and concentration parameter $c$ are expressed
in terms of a bivariate distribution function ${d^2 n(M,c;z)/dM dc}
= {dn/dM}(M;z) P(c|M;z)$. The conditional probability distribution
function of a halo having concentration $c$ for a given mass $M$ in
numerical simulations was found to be well approximated by a
log-normal distribution. $P(c|M;z) dc = {}\exp \left [ -{(\log c -
\log \bar c) }/ 2\sigma_{\log c} \right] {dc/ (c \; ln(10))}$ with $\bar
c(M,z) = {c_0/ 1+z} \left [ {M /M_{\star}(z=0)} \right
]^{-\alpha_c}$. Here $M_{\star}$ is the mass scale at which
$\nu(M_{\star};z)=1$ and $c_0=9$ and $\alpha_c=0.13$ and
$\sigma_{log c}=0.14$. More complicated models are possible where a
mass dependent $\sigma_{log c}$ can be incorporated. For more discussion
on equilibrium baryonic gas profile and its relation with the 
underlying halo profile see \cite{CBS05}. We have considered the halo mass
range $10^{11}M_{\sun}$\;-\;$10^{16}_{\sun}$ in our calculations.
\begin{figure}
\begin{center}
{\epsfxsize=15 cm \epsfysize=5 cm {\epsfbox[27 517 587 709]{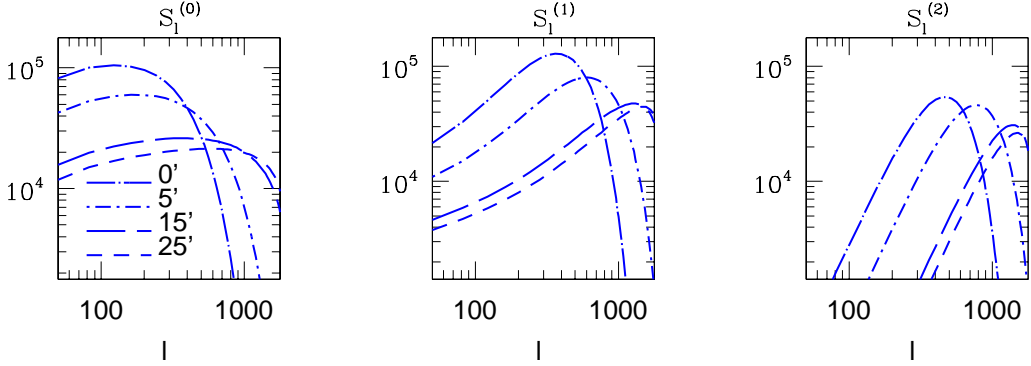}}}
\end{center}
\caption{The skew-spectra are plotted as a function of the harmonic $l$. From left to right
the plots correspond to $S_l^{(0)}$, $S_l^{(1)}$ and $S_l^{(2)}$ respectively. We consider
four different beams with FWHM = $0'$, $5'$, $15'$ and $25'$ respectively as indicated.
These results correspond to an ideal all-sky noise fee experiment. A perturbative calculation
which assumes redshift dependent linear biasing was used.}
\label{fig:skew_spec2c}
\end{figure}
\begin{figure}
\begin{center}
{\epsfxsize=15 cm \epsfysize=5 cm {\epsfbox[27 517 587 709]{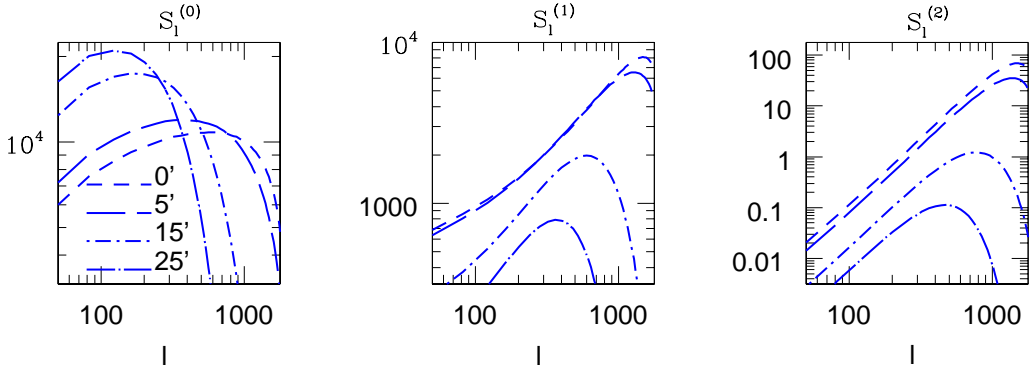}}}
\end{center}
\caption{The skew-spectra are plotted as a function of the harmonic $l$. From left to right
the plots correspond to $S^{(0)}$, $S^{(1)}$ and $S^{(2)}$ respectively. We consider
 four different beams with FWHM = $0'$, $5'$, $15'$ and $25'$ respectively as indicated.
These results include Planck-type noise.}
\label{fig:skew_spec2d}
\end{figure}
\section{Minkowski Functionals}
\label{sec:MF_intro} In this section we will review the basics of
the Minkowski Functional (MFs) in brief.  The MFs are morphological
descriptors well known in cosmology. Put simply, morphological
properties are the properties that remain invariant under rotation
and translation; see e.g. \cite{Hadwiger59} for a more mathematical
discussion. We will concentrate on a set of three MFs defined in 2D
that are defined over an excursion set $\Sigma$ with a boundary
$\partial\Sigma$ for a given threshold $\nu$. Following the notation
of \cite{Hk08} we define:
\be
V_0(\nu) = \int_{\Sigma} da; \quad V_1(\nu) = {1 \over 4}\int_{\partial\Sigma} dl;
\quad V_2(\nu) = {1 \over 2\pi}\int_{\partial \Sigma} {\cal K} dl.
\label{eq:mf}
\ee
Here $\cal K$ denotes the curvature along the boundary $\partial \Sigma$.
The three MFs $V_0(\nu)$, $V_1(\nu)$ and $V_2(\nu)$,
correspond to the area of the excursion set $\Sigma$, the length of its boundary $\partial\Sigma$ as well as the
integral of curvature $\cal K$ along its boundary which is also related to the genus and hence
the Euler characteristics. The MFs
can be decomposed into two different contributions Gaussian  $V^G_k(\nu)$ and non-Gaussian $\delta V_k(\nu)$
i.e.  $V_k(\nu) = V_k^G(\nu) + \delta V_k(\nu)$ with $\nu= \myf/\sigma_0$ and $\sigma^2_0=\la \myf^2 \ra$.
We non-Gaussian contribution i.e. $\delta V_k(\nu)$.
We will further separate out an amplitude $A$ in the expressions of both of these contributions which depend only on the power spectrum of the  perturbation
through $\sigma_0$ and $\sigma_1$ (see e.g. \cite{Hk08}):
\ben
&& V^G_k(\nu) = A \exp \left ( -{\nu^2 \over 2}\right )  H_{k-1}; \quad\quad
\delta V_k(\nu) = A \exp \left ( -{\nu^2 \over 2}\right )
\left [ \delta V_k^{(2)}(\nu)\sigma_0 + \delta V_k^{(3)}(\nu)\sigma_0^2 + \delta V_k^{(4)}(\nu)\sigma_0^3 + \cdots \right ]  \\
&& \delta V_k^{(2)}(\nu) = \left [ \left \{  {1 \over 6} S^{(0)} H_{k+2}(\nu) + {k \over 3} S^{(1)} H_k(\nu) + {k(k-1) \over 6} S^{(2)} H_{k-2}(\nu)\right \} \right ];
\quad\quad
A = {1 \over (2\pi)^{(k+1)/2}} {\omega_2 \over \omega_{2-k}\omega_k}\left ( \sigma_1 \over \sqrt 2 \sigma_0 \right )^k.
\label{eq:v_k}
\een
The constant $\omega_k$ introduced above is the volume of the unit
sphere in $k$ dimensions. $w_k = {\pi^{k/2}/ \Gamma(k/2+1)}$ in 2D
we will only need $\omega_0=1$, $\omega_1=2$ and $\omega_2 =\pi$.
The Hermite polynoimals $H_k(\nu)$ appear in the expression of MFs
for weakly non-Gaussian random field. The lowest-order departure
from Gaussianity $\delta V_k^{(2)}(\nu)$ is encoded in three
different generalized skewness parameters  $S^{(0)}, S^{(1)},
S^{(2)}$. The next to leading order correction terms $\delta
V_k^{(2)}(\nu)$ are typically neglected as they are order of
magnitude smaller. Just as the generalized three skewness parameters
they can be constructed from generalization of kurtosis. All three
skewness parameters can be constructed from the bispectrum using
different weights to sample different modes. In our next section we
will discuss these generalized skewness parameters and introduce the
power spectra associated with them. We will denote the power spectra
with $S^{(0)}_l, S^{(1)}_l$ and $S^{(2)}_l$. These will carry more
information than their one-point counterparts the generalized
skewness parameters.

Estimation of MFs typically is done directly by employing a
grid-based approach in real space, so that the integrals introduced
in Eq.(\ref{eq:mf}) are estimated on a discrete set of points. Here
we will show that the computation of $S^{(i)}_l$ introduced in
Eq.(\ref{eq:v_k}) can also be done in harmonic space. This will be
particularly suitable for near all-sky surveys where galactic masks
and point source masks can be incorporated in the analysis
relatively easily. This will allow a cross-comparison of both
methods for any systematics as well as residuals from component
separation. We will deal with near all-sky surveys but flat-sky
results can also be recovered using standard procedure which will be
suitable for smaller patch-sky surveys.

The Fig.\ref{fig:one} shows the parameters $S^{(0)}$, $S^{(1)}$ and
$S^{(2)}$ as a function of FWHM assuming a Gaussian beam. We have
assumed a halo model for computation of these generalized skewness
parameters. The parameters defined in Eq. (\ref{eq:v_k}), $\sigma_0$
and $\sigma_1$, appear in the perturbative expansion of the MFs as
well as in their normalization. These parameters depend on the level of noise.
\section{Generalised Skew-Spectra and MFs}
\label{sec:MF}
The skew-spectra are cubic statistics that are constructed by cross-correlating two different fields.
One of the field used is a composite field typically a product of two maps either in its original form or constructed
by means of relevant differential operations. The second field will typically
be a single field but may be constructed by using differential operators on the original field.
All of these three skewness contribute to the three MFs that we will consider in 2D.
The first of the skew-spectra was studied by \citep{Cooray01}
and was later generalized by \cite{MuHe10} and is related to commonly used skewness.
The skewness in this case is constructed by cross-correlating the
squared map $[\myf^2(\oh)]$ with the original map $[\myf(\oh)]$.  The
second skew-spectrum is constructed by cross-correlating the squared map $[\myf^2(\oh)]$ against $[\nabla^2\myf(\oh)]$.
Analogously the third skew-spectrum represents the cross-spectra that can be constructed
using $[\nabla \myf(\oh)\cdot\nabla \myf(\oh)]$ and $[\nabla^2 \myf(\oh)]$ maps.
\ben
&& S_l^{(0)} \equiv {1 \over 12 \pi \sigma_0^4}S_l^{(\myf^2,\myf)} \equiv {1 \over 12 \pi \sigma_0^4}{1 \over 2l+1}\sum_m
{\rm Real}([\myf]_{lm}[\myf^2]^*_{lm})  ={1 \over 12 \pi \sigma_0^4} \sum_{l_1l_2} \myB_{ll_1l_2}J_{ll_1l_2}W_{l}W_{l_1}W_{l_2}  \\
&& S_l^{(1)} \equiv {1 \over 16 \pi \sigma_0^2\sigma_1^2} S_l^{(\myf^2,\nabla^2 \myf)}
\equiv {1 \over 16 \pi \sigma_0^2\sigma_1^2}{1 \over 2l+1}\sum_m {\rm Real}([\nabla^2 \myf]_{lm}[\myf^2]^*_{lm}) \nn \\
&&\quad\quad = {1 \over 16 \pi \sigma_0^2\sigma_1^2} \sum_{l_i} \Big [{l(l+1)+ l_1(l_1+1)+ l_2(l_2+1)} \Big ] \myB_{ll_1l_2}J_{ll_1l_2}
W_{l}W_{l_1}W_{l_2}  \\
&& S_l^{(2)} \equiv {1 \over 8 \pi \sigma_1^4}S_l^{(\nabla \myf\cdot\nabla \myf, \nabla^2\myf)} \equiv {1 \over 8 \pi \sigma_1^4} {1 \over 2l+1}\sum_m
{\rm Real}([\nabla \myf \cdot \nabla \myf]_{lm}[\nabla^2 \myf]^*_{lm}) \nn \\
&&\quad\quad ={1 \over 8 \pi \sigma_1^4} \sum_{l_i}
{}\Big [ [l(l+1)+l_1(l_1+1) - l_2(l_2+1)]l_2(l_2+1) + {\rm cyc.perm.} \Big ]
\myB_{ll_1l_2}J_{ll_1l_2}W_{l}W_{l_1}W_{l_2}\\
&& J_{l_1l_2l_3} \equiv {I_{l_1l_2l_3} \over 2l_3+1} = \sqrt{(2l_1+1)(2l_2+1) \over (2l_3+1) 4 \pi }\left ( \begin{array}{ c c c }
     l_1 & l_2 & l_3 \\
     0 & 0 & 0
  \end{array} \right). \\
&& S^{(i)} = \sum_{l}(2l+1)S^{(i)}_l\\
&& \sigma_j^2 = {1 \over 4\pi}\sum_l (2l+1)[l(l+1)]^j \myC_l W_l^2
\label{skew_spectra}
\een
This set of equations constitutes one of the main results of this
paper. The matrices here denote the Wigner $3j$ symbols \cite{Ed68},
$W_l$ represents the smoothing window which can be, e.g.,  top-hat
or Gaussian or compensated. Each of these spectra probes the same
bispectrum $B_{ll_1l_2}$ but with different weights for individual
triplets of modes  $(l,l_1l_2)$ specifying a triangle in the
harmonic domain. The skew-spectrum sums over all possible
configuration of the bispectrum keeping one of its side $l$ fixed.
For each individual choice of $l$ we can compute the skew-spectra
$S_l^{(i)}$. The extraction of skew-spectra from data is relatively
straight forward. It consists of construction of the relevant maps
in real space either by algebraic or differential operation and then
cross-correlating them in the multipole domain. Issues related to
mask and noise will be dealt with in later sections. We will show
that even in the presence of mask the computed skew spectra can be
inverted to give a unbiased estimate of all-sky skew-spectra.
Presence of noise will only affect the scatter. We have explicitly
displayed the experimental beam $b_l$ in all our expressions.

To derive the above expressions, we first express the spherical
harmonic expansion of fields the $[\nabla^2 \myf(\oh)]$, $[\nabla
\myf(\oh) \cdot \nabla \myf(\oh)]$ and $[\myf^2(\oh)]$ in terms of
the harmonics of the original fields $\myf_{lm}$. These expressions
involve the $3j$ functions as well as factors that depend on various
$l_i$ dependent weight factors.
\ben
&& [\nabla^2 \myf(\oh)]_{lm} = \int ~d\oh~ Y^*_{lm}(\oh)~ [\nabla^2 \myf(\oh)] = -l(l+1)\myf_{lm} \nn \\
&&[\myf^2(\oh)]_{lm} = \int~d\oh~Y^*_{lm}(\oh)~[\myf^2(\oh)] = \sum_{l_im_i} (-1)^m\myf_{l_1m_1} \myf_{l_2m_2}I_{l_1l_2l}
\left ( \begin{array}{ c c c }
     l_1 & l_2 & l\\
     m_1 & m_2 & -m
  \end{array} \right). \nn \\
&& [\nabla \myf(\oh) \cdot \nabla \myf(\oh)]_{lm}  = \int d\oh Y^*_{lm}(\oh)[\nabla \myf(\oh) \cdot \nabla \myf(\oh)] =
\sum_{l_im_i}  \myf_{l_1m_1} \myf_{l_2m_2} \int~d\oh~Y^*_{lm}(\oh)~ [\nabla Y_{l_1m_2}(\oh) \cdot \nabla Y_{l_2m_2}(\oh)] \\
&& \quad\quad\quad  = {1 \over 3}  \sum_{l_im_i}[l_1(l_1+1)+l_2(l_2+1)-l(l+1)] \int d\oh Y^*_{lm}(\oh) Y_{l_1m_1}(\oh)Y_{l_2m_2}(\oh) \nn \\
&& \quad\quad\quad  = {1 \over 3} \sum_{l_im_i}(-1)^m[l_1(l_1+1)+l_2(l_2+1)-l(l+1)]
\myf_{l_1m_1} \myf_{l_2m_2} I_{l_1l_2l}
\left ( \begin{array}{ c c c }
     l_1 & l_2 & l \\
     m_1 & m_2 & -m
  \end{array} \right).
\label{eq.harmonics}
\een
\n
We can define the power spectrum associated with the MFs through the following third order expression:
\be
V_k^{(3)} = \sum_l[V_k]_l(2l+1) = {1 \over 6 } \sum_l (2l+1) \left \{ S^{(0)}_l H(\nu) + {k \over 3 } S^{(1)}_l H(\nu) +
{k(k-1) \over 6 } S^{(2)}_l H_{k-2}(\nu) + \cdots \right \}.
\ee
\n The ``three-skewness'' defines the triplets of Minkowski
Functionals (MFs). At the level of two-point statistics, in the
harmonic domain we have three power-spectra associated with
Minkowski-Functional $V_k^{(3)}$ that depend on the three
skew-spectra we have defined. We will show later in this paper that
the fourth order correction terms also have a similar form, but with
an additional monopole contribution that can be computed from the
lower order one-point terms such as the three skewness defined here.
The result presented here is important and implies that we can study
the contributions to each of the MFs $v_k(\nu)$ as a function of
harmonic mode $l$. This is especially significant result as various
form of non-Gaussianity will have different $l$ dependence and hence
they can potentially be distinguished. The ordinary MFs add
contributions from all individual $l$ modes and hence have less
power in differentiating various contributing sources of
non-Gaussianity. This is one of main motivations for extending the
concept of MFs (single numbers) to one-dimensional objects similar
to power spectrum.
It is worth pointing out that the skewness, along with the
generalized skewness parameters, are typically less sensitive to the
background cosmology. They are more sensitive to the underlying
model of non-Gaussianity. The main dependence on cosmology typically
results from the normalization coefficients such as $\sigma_0$ and
$\sigma_1$ which are determined using the power spectrum of tSZ i.e.
$\Theta$.

In Figure-\ref{fig:skew_spec2a} we plot the three skew spectra for a range of 
Gaussian beam with FWHM $= 0'- 25'$. We consider an ideal all-sky experimental
set up without any instrumental noise. The analytical results are
computed using a halo model prediction for the bispectrum. The corresponding
skew-spectra that include Planck type noise are displayed in Figure-\ref{fig:skew_spec2b}.
As expected the effect of noise is more pronounced at smaller angular scales.
In  Figure-\ref{fig:skew_spec2c} ideal noise free all-sky results are plotted for
the diffuse component. The bispectrum in this case is constructed using
the perturbative approach. The Figure-\ref{fig:skew_spec2d} shows the 
corresponding results for a Planck type all-sky experiments. Notice that
for near all-sky experiments the signal-to-noise will degrade as $f_{sky}$. 
All results shown are for $f_{sky}=1$.
\section{Estimators and Their Scatter}
\label{sec:scatter}
\n As we have noted before, the estimators for the skew-spectra can
be most easily computed by cross-correlating maps in the harmonic
domain. These maps are constructed in real space by applying various
derivative operators. In the presence of mask the recovered
skew-spectra will depend on the mask. The presence of mask typically
introduces a mode-mode coupling. The approach we adopt here to
reconstruct the unbiased power spectra in the presence of mask
exploits the so-called Pseudo-$\myC_l$  \citep{Hiv}, which involves
expressing the observed power spectra $\myC_l$ in the presence of a
mask as a linear combination of unbiased all-sky power spectra.

The three different generalized skew spectra that we have introduced
here can be thought as cross-spectra of relevant fields. We denote
these generic fields by $A$ and $B$ and will denote the generic
skew-spectra as $S_l^{A,B}$. The skew-spectra that is recovered in
the presence of mask will be represented as $\tilde S_l^{A,B}$ and
the unbiased estimator will be denoted by $\hat S_l^{A,B}$. The
skew-spectra recovered in the presence of mask $\tilde S_l^{A,B}$
will be biased. However to construct an unbiased estimator  $\hat
S_l^{A,B}$ for the skew-spectra the following procedure is
sufficient. The derivation follows the same arguments as detailed in
\cite{Mu09} and will not be reproduced here.
\be
\tilde S_l^{A,B} = {1 \over 2l +1} \sum_{m} \tilde A_{lm} \tilde B^*_{lm}; \quad
\tilde S_l^{A,B} = \sum_{l'} M_{ll'} S_l^{A,B}; \quad M_{ll'} = {1 \over 2l + 1}\sum_{l'l''} I^2_{ll'l''} |w_{l''}|^2; \quad \left \{ A,B \right \} \in \left \{\myf,\myf^2, (\nabla \myf\cdot \nabla \myf), \nabla^2 \myf \right\}.  \\
\label{eq:alm_est}
\ee
\n
In this notation we can write the skewness parameters defined previously as:
\be
S^{(0)} \equiv {\rm X}_{(0)}S_l^{(\myf^2,\myf)};\quad %= {1 \over 12 \pi \sigma_0^4}S_l^{(\myf^2,\myf)} ; \quad
S^{(1)} \equiv {\rm X}_{(1)}S_l^{(\myf^2,\nabla^2\myf)};\quad %= {1 \over 16 \pi \sigma_0^2 \sigma_1^2}S_l^{(\myf^2, \nabla^2\myf)}; \quad
S^{(2)} \equiv {\rm X}_{(2)}S_l^{(\nabla \myf\cdot\nabla \myf, \nabla^2\myf)}
%= {1 \over 8 \pi \sigma_1^4} S_l^{(\nabla \myf\cdot\nabla \myf, \nabla^2\myf)}
\quad\quad \{{\rm X}_{(0)}, {\rm X}_{(1)}, {\rm X}_{(2)}\} \equiv
\left \{ {1 \over 12 \pi \sigma_0^4}, {1 \over 16 \pi \sigma_0^2 \sigma_1^2},{1 \over 8 \pi \sigma_1^4}\right \}
\ee
\n
The mode-mode coupling matrix $M$ is constructed from the power spectra of the mask $w_{l''}$ and used
for estimation of unbiased skew-spectra $\hat S_{l'}^{A,B}$. Typically the mask consists of bright stars
and saturated spikes where no lensing measurements can be performed. The results that we present here
are generic. The estimator thus constructed is an unbiased
estimator. The computation of covariance of the the scatter in the estimates can be computed using analytical methods,
thereby avoiding the need of expensive Monte-Carlo simulations. The scatter or covariance of the unbiased
estimates $\la \delta \hat S_l^{A,B} \delta \hat S_{l'}^{A,B} \ra$ is related to that of the direct estimates
$\la \delta \tilde S_l^{A,B} \delta \tilde S_{l'}^{A,B} \ra$ from the masked sky by a similarity transformation.
The transformation is given by the same mode coupling matrix $M$.
\be
\hat S_l^{A,B} = \sum_{l'} [M^{-1}]_{ll'} \tilde S_{l'}^{A,B}; \quad
\la \delta \hat S_l^{A,B} \delta \hat S_{l'}^{A,B} \ra =  \sum_{LL'}M^{-1}_{lL} \la \delta \tilde S_{L}^{A,B} \delta \tilde S_{L'}^{A.B} \ra M^{-1}_{L'l'};
\quad \langle \hat S_l^{A,B} \rangle = S_l^{A,B}; \quad \left \{ A,B \right \} \in \left \{\myf,\myf^2, (\nabla \myf\cdot \nabla \myf), \nabla^2 \myf \right\}.  \\
\label{eq:auto_cov} \ee \n The power spectra associated with the MFs
are linear combinations of the skew-spectra (see Eq.(\ref{eq:v_k})).
In our approach the power spectra associated with the MFs are
secondary and can be constructed using the skew-spectra that are
estimated directly from the data.

The construction of an estimator is incomplete without an estimate
of its variance. The variance or the scatter in certain situation is
computed using Monte-Carlo (MC) simulations which are
computationally expensive. In our approach, it is possible to
compute the covariance of our estimates of various $S_l$s, i.e.
${\langle \delta S_l \delta S_{l'} \rangle}$ under the same
simplifying assumptions that higher order correlation functions can
be approximated as Gaussian. This allows us to express the error
covariance in terms of the relevant power spectra. The generic
expression can be written as: \be [\hat V_k^{(2)}]_l =  \sum_{l'}
[M^{-1}]_{ll'} [\tilde V_k^{(2)}]_l; \quad \quad \la \delta \hat
V_k^{(2)} \delta \hat V_{k'}^{(2)} \ra = \sum_{LL'}M^{-1}_{lL} \la
\delta [\tilde V_{k}^{(2)}]_l \delta [\tilde V_{k'}^{(2)}]_{l'} \ra
M^{-1}_{L'l'} \ee \n We would like to point out here that, in case
of limited sky coverage, it may not be possible to estimate the
skew-spectra mode by mode as the mode coupling matrix may become
singular and a broad binning of the spectra may be required.
\begin{figure}
\begin{center}
{\epsfxsize=15 cm \epsfysize=5 cm {\epsfbox[27 517 587 709]{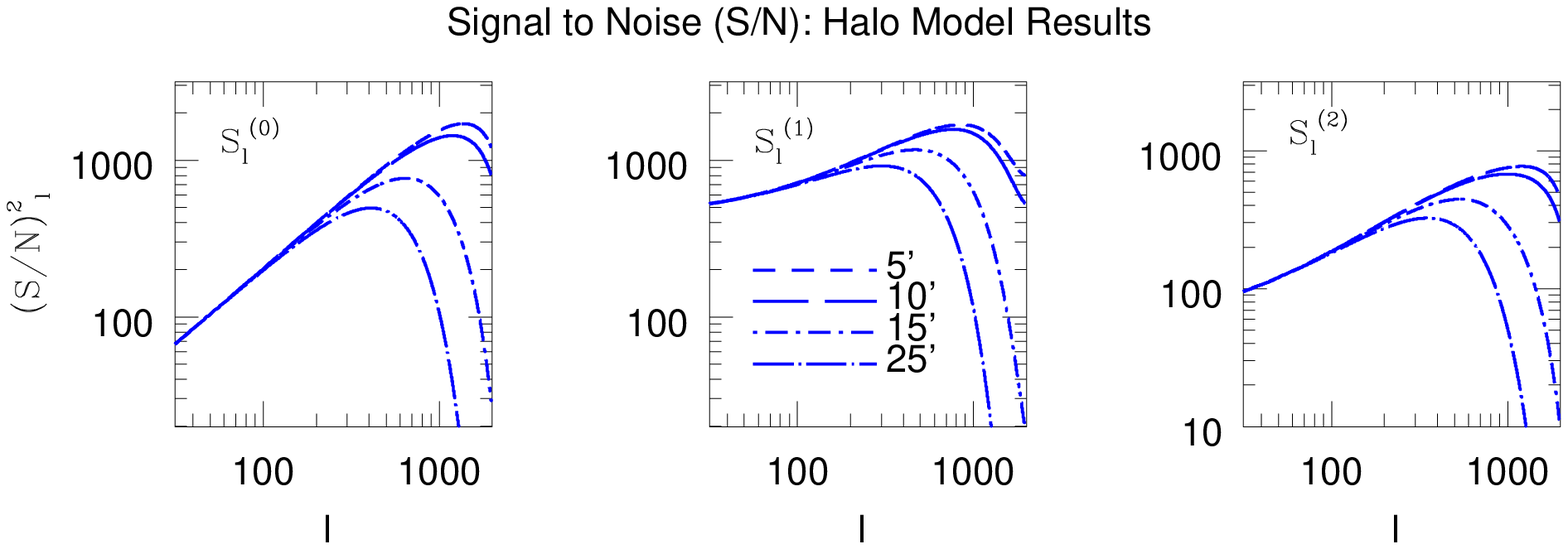}}}
\end{center}
\caption{The signal-to-noise (S/N) squared for various skew-spectra $S_l^{(0)}$ (left panel), $S_l^{(1)}$ (middle panel)
and $S_l^{(2)}$ (right panel) are plotted as a function of angular harmonics $l$. We choose four different Gaussian beams with FWHM
$\theta_s = 0', 5', 15, 25'$. The higher resolution (smaller FWHM) reaches higher signal-to-noise.
The plots correspond to an ideal experiment with all-sky coverage and no instrumental noise. The expressions used for computation of the signal-to-noise are given in Eq.(\ref{s2n0})-Eq.(\ref{s2n2}). The
results correspond to an ideal all-sky $f_{sky}=1$ {\it no-noise} setup. For near all-sky esperiments the
signal to noise will scale directly with $f_{sky}$.}
\label{fig:s2na}
\end{figure}
\beqa
 && \langle [\delta S_l^{X,Y}]\delta S_{l'}^{X,Y}] \rangle =
f^{-1}_{sky}{1 \over 2l+1} \left [ \myC_l^{X,X} \myC_{l'}^{Y,Y}
+ [S_l^{X,Y}]^2  \right ]\delta_{ll'}; \quad\quad \{X,Y\} \in \{\Theta, \Theta^2, \nabla \Theta(\oh) \cdot \nabla \Theta(\oh),
\nabla^2\Theta(\oh) \}.
\label{eq:error_cov}
\eeqa
Here the fraction of sky covered by the survey is denoted by $f_{sky}$.
The expressions for the skew-spectra are quoted in  $S_l^{(\Theta^2,\Theta)}$,$S_l^{(\kappa^2,\nabla^2\kappa)}$ and
$S_l^{(\nabla\Theta\cdot\nabla\Theta,\nabla^2\Theta)}$  are given in Eq.(\ref{skew_spectra}). The expressions
for covariance also depend on a set of
power spectra  i.e. $S_l^{(\Theta^2,\Theta^2)}$, $S_l^{(\nabla^2\Theta,\nabla^2\Theta)}$,
 $S_l^{(\nabla\Theta \cdot\nabla\Theta,\nabla^2\Theta)}$ and $S_l^{\Theta,\Theta}$. These are given by the following expression:
\ben
&& \myC_l^{\nabla\cdot\nabla,\nabla\cdot\nabla} = \sum_{l_1l_2} \myC^{t}_{l_1} \myC^{t}_{l_2}  [l_1(l_1+1)+l_2(l_2+1)-l(l+1)]^2  I^2_{ll_1l_2}; \quad
\myC_{l}^{[\myf^2, \myf^2]} = \sum_{l_1l_2} \myC^{t}_{l_1} \myC^{t}_{l_2} I^2_{ll_1l_2}; \quad
 \myC_{l}^{[\nabla^2 \myf, \nabla^2 \myf]} = l^2(l+1)^2\myC^{t}_l
\een
\n
Here $\myC_l^t$ includes signal and noise power spectra $\myC^t_l=\myC^S_lb^2_l + \myC^N_l$. The experimental
beam is denoted by $b_l$. Using these equations it is
possible to compute the scatter in various skew-spectra. The scatter will depend on the skew-spectra
as well as various cross-spectra of various combination of product fields constructed from the original tSZ maps.
\ben
&&
\langle \delta S_l^{[\Theta^2,\Theta]} \delta S_l^{[\Theta^2,\Theta]} \rangle = f^{-1}_{\rm sky} {1 \over 2l + 1}
\left [ \myC_l^{[\Theta^2,\Theta^2]}\myC_l^{[\Theta,\Theta]} + [S_l^{[\Theta^2,\Theta]}]^2 \right ] \\
&& \langle \delta S_l^{[\Theta^2,\nabla^2 \Theta]} \delta S_l^{[\Theta^2,\nabla^2 \Theta]} \rangle = f^{-1}_{\rm sky} {1 \over 2l + 1}
\left [ \myC_l^{[\Theta^2,\Theta^2]}\myC_l^{[\nabla\cdot\nabla,\nabla\cdot\nabla]} +
[ S_l^{[\Theta^2,\nabla^2\Theta]} ]^2 \right ] \\
&&  \langle \delta
S_l^{[\nabla\Theta\cdot\nabla\Theta,\nabla^2\Theta]} \delta
S_l^{[\nabla\Theta\cdot\nabla\Theta,\nabla^2\Theta]} \rangle =
f^{-1}_{\rm sky} {1 \over 2l + 1}\left [
\myC_l^{[\nabla^2\Theta,\nabla^2\Theta]}\myC_l^{[\nabla\cdot\nabla,\nabla\cdot\nabla]}
+ [S_l^{[\nabla\Theta\cdot\nabla\Theta,\nabla^2\Theta]}]^2 \right ]
\een \n The cumulative signal to noise up to a given $l_{\rm max}$
using these expression for estimators $S^{(0)}$ can now be expressed
as: \ben && \left [\left ( {S / N } \right )^{(0)}_{l_{\rm
max}}\right ]^2 = f_{sky}\sum_{l=2}^{l_{\rm max}} (2l+1)
 (S_l^{[\Theta^2,\Theta]})^2 / \left [ {\rm X}_{(0)}^2\myC_l^{[\Theta^2,\Theta^2]}\myC_l^{[\Theta,\Theta]} + (S_l^{[\Theta^2,\Theta]})^2 \right ]  \label{s2n0}\\
&& \left [\left ( {S / N } \right )^{(1)}_{l_{\rm max}}\right ]^2 =
f_{sky}\sum_{l=2}^{l_{\rm max}} (2l+1)
 {(S_l^{[\Theta^2,\nabla^2\Theta]})^2 / \left [  {\rm X}_{(1)}^2\myC_l^{[\Theta^2,\Theta^2]}\myC_l^{[\nabla^2,\nabla^2]} +
[ S_l^{[\Theta^2,\nabla^2\Theta]} ]^2 \right ] } \label{s2n1}\\
&& \left [\left ( {S / N } \right )^{(2)}_{l_{\rm max}}\right ]^2 =
f_{sky}\sum_{l=2}^{l_{\rm max}} (2l+1)
 (S_l^{[\nabla\Theta\cdot\nabla\Theta,\nabla^2\Theta]})^2 / \left [ {\rm X}_{(2)}^2\myC_l^{[\nabla^2\Theta,\nabla^2\Theta]}\myC_l^{[\nabla\cdot\nabla,\nabla\cdot\nabla]} +
[S_l^{[\nabla\Theta\cdot\nabla\Theta,\nabla^2\Theta]}]^2 \right ]
\label{s2n2} \een \n For the study of the tSZ effect we find that a
robust determination of the generalised skew-spectra is possible for
individual modes for almost the entire range of $l$ values we have
probed. This is important in differentiating them from various other
sources of non-Gaussianity. The skew-spectra are integrated
statistics, meaning that their value at a given $l$ depends on the
entire range of $l$ values probed. These results can also be
extended to take into account the cross-correlation among various
skew-spectra extracted from the same data. \ben && \langle \delta
S_l^{[\Theta^2,\Theta]} \delta S_l^{[\Theta^2,\nabla^2\Theta]}
\rangle = f^{-1}_{\rm sky} {1 \over 2l + 1} \left [
\myC_l^{[\Theta^2,\Theta^2]}\myC_l^{[\Theta,\nabla^2\Theta]} +
S_l^{[\Theta^2,\nabla^2\Theta]}
S_l^{[\Theta,\Theta^2]}\right ] \label{eq:10_cross} \\
&& \langle \delta S_l^{[\Theta^2,\Theta]} \delta S_l^{[\nabla\Theta\cdot\nabla\Theta,\nabla^2\Theta]} \rangle = f^{-1}_{\rm sky} {1 \over 2l + 1}
\left [ S_l^{[\Theta^2,\nabla^2\Theta]}\myC_l^{[\Theta,\nabla\Theta\cdot\nabla\Theta]} +  \myC_l^{(\Theta^2,\nabla\Theta\cdot\nabla\Theta)}
 \myC_l^{(\Theta,\nabla^2\Theta)}   \right ] \\
&& \langle \delta S_l^{[\Theta^2,\nabla^2\Theta]} \delta S_l^{[\nabla\Theta\cdot\nabla\Theta,\nabla^2\Theta]} \rangle =
f^{-1}_{\rm sky} {1 \over 2l + 1}
\left [  \myC_l^{(\Theta^2,\nabla\Theta\cdot\nabla\Theta)}
 \myC_l^{(\nabla^2\Theta,\nabla^2\Theta)} + S_l^{[\Theta^2,\nabla^2\Theta]}S_l^{[\nabla^2\Theta,\nabla\Theta\cdot\nabla\Theta]} \right ]
\een This discussion involves the lowest order departure from
Gaussianity in MFs using a third order statistic, namely the
bispectrum. The next to leading descriptions are characterized by
the trispectrum which is a fourth order statistics. It is possible
to extend the definition of skew-spectra to the case of kurt-spectra
or the power spectrum associated with tri-spectra. The power spectra
associated with the Minkowski Functionals can be defined completely
up to fourth order using the skew- and the kurt-spectra. However the
corrections to leading order statistics from kurt-spectra are
sub-dominant and leading order terms are enough to study the
departure from Gaussinaity. In any case it is nevertheless straight
forward to implement an estimator which which will estimate the
power-spectrum associated with the MFs from a noisy data by
including both third order and fourth order statistics is relatively
straight forward. The issues has been dealt with in detail in
\citep{MSC10a} in the context of CMB sky.

In addition to the three generalized skew-spectra that define the
MFs at lowest order in non-Gaussianity, it is possible also to
construct additional skew-spectra that work with different set of
weights. In principle an arbitrary number of such skew-spectra can
be constructed though they will not have direct links with the
morphological properties that we have focused on, in this paper they
can still be used as a source of independent information on the
bispectrum and can be used in principle to separate sources of
non-Gaussianity, whether it be  primordial or induced by late-time
gravitational effects.
\begin{figure}
\begin{center}
{\epsfxsize=15 cm \epsfysize=5 cm {\epsfbox[30 534 587 709]{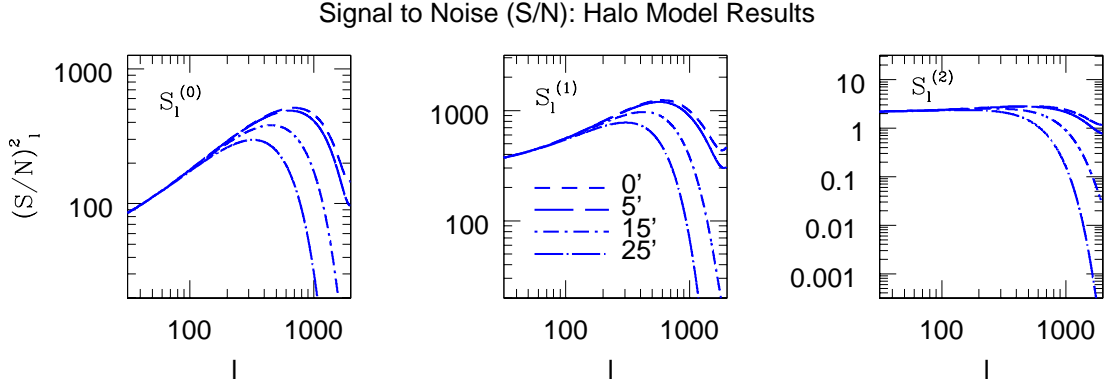}}}
\end{center}
\caption{Same as previous figure but with Planck noise added. Inclusion of noise degrades the signal-to-noise.
The effect is most prominent for $S_l^{(2)}$ which is related to the fact that $S_l^{(2)}$ gives more
weights to smaller scales than other estimates and hence more affected by the noise. }
\label{fig:s2nb}
\end{figure}
\begin{figure}
\begin{center}
{\epsfxsize=15 cm \epsfysize=5 cm {\epsfbox[30 534 587 709]{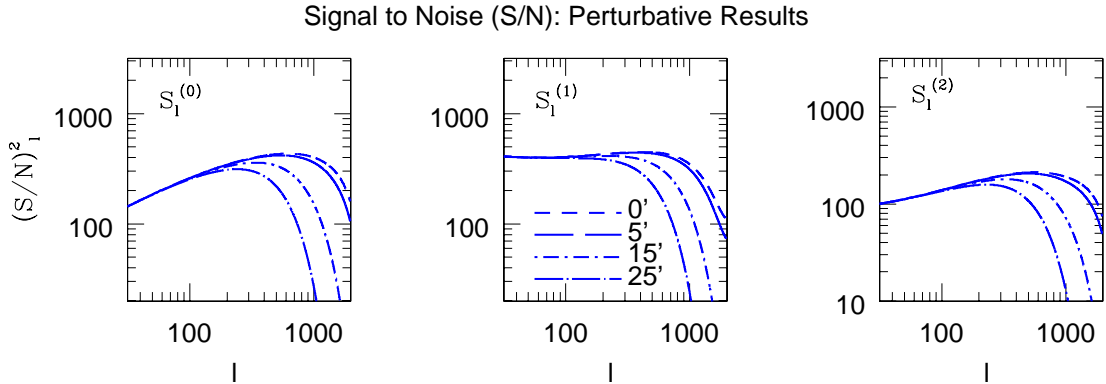}}}
\end{center}
\caption{Same as Figure-\ref{fig:s2nb} but for perturbative calculations. The perturbative bispectrum
has less power at higher $l$ compared to halo model calculations.} 
\label{fig:s2nc}
\end{figure}
\begin{figure}
\begin{center}
{\epsfxsize=15 cm \epsfysize=5 cm {\epsfbox[30 534 587 709]{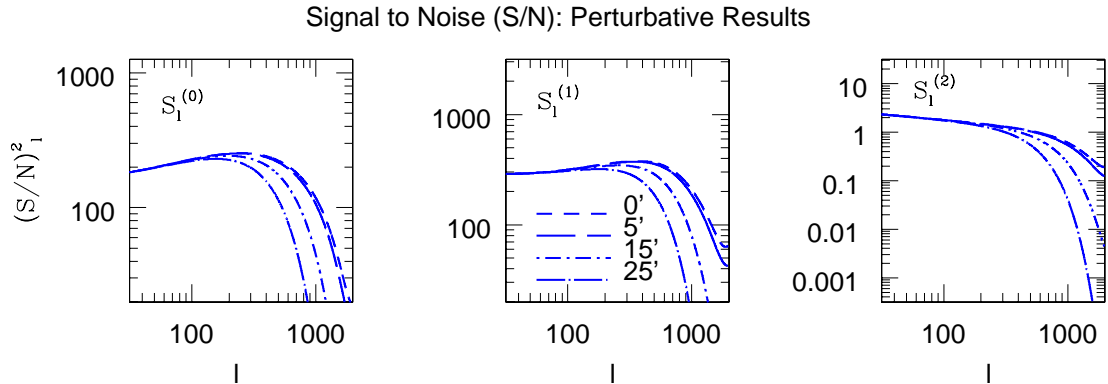}}}
\end{center}
\caption{These results correspond to perturbative analysis but with instrumental noise for a
Planck-type experiment included.}
\label{fig:s2nd}
\end{figure}
In Figure-\ref{fig:s2na} we have plotted the signal to noise
for the three skew-spectra for different choice of the FWHM as indicated for an ideal experiment.
The analytical results for the underlying bispectrum is obtained using halo model prescription. 
The left, middle and right panel correspond to $S_l^{(0)}$, $S_l^{(1)}$ and $S_l^{(2)}$ respectively.
In Figure-\ref{fig:s2nb} the signal-to-noise for an all-sky experiment with Planck type experiment
is shown. As expected the
with smaller beam size will have higher signal-to-noise. In Figure-\ref{fig:s2nc} results for Planck type
experiment is shown but using perturbative calculations and finally we show perturbative
results that include noise in Figure-\ref{fig:s2nc}. 
The skew-spectra $S_l^{(2)}$ has higher signal-to-noise among the three
skew-spectra. The very high signal to noise will allow an accurate determination of all the
three spectra especially $S_l^{(0)}$ and $S_l^{(1)}$ for the entire range of $l$ values probed.
\section{Conclusion}
\label{sec:conclu} In this study we have examined the prospects for
extracting  non-Gaussianity statistics from CMB
surveys using MFs to probe the non-Gaussianity that originates from
the tSZ effect. The tSZ effect is associated with the hot gas in
large-scale structure that is detectable by multi-frequency
experiments, in particular Planck.. When compared with the CMB
temperature anisotropies the tSZ effect has a distinct spectral
signature which means it can be effectively separated from the
primary CMB contributions. This will provide an unique opportunity
to probe tSZ effect using frequency-cleaned data in the very near
future. In the past, the statistical analysis of the frequency
cleaned tSZ maps has so far been mainly focused on lower order
statistics. Morphological studies involving the MFs are
complementary to the lower order statistics and carry independent
information. In this paper we have generalized the study of MFs that
are typically used in various other context in cosmology to the case
of tSZ maps.

As previously mentioned, the tSZ effect traces the fluctuations
associated with the large scale distribution of the baryonic gas.
The generation of pressure fluctuation in the virialized dark matter
halos can be modeled by assuming them to be in hydrostatic
equilibrium with the dark matter distribution within the halo. The
shock heated gas typically correspond to over-densities in excess of
$\delta \ge 200$. The unshocked photoionized baryonic gas typically
traces the large scale distribution of the dark matter distribution.
The typical over-density in unshocked photoionised baryons is
typically  $ \delta \le 10$.

We have modelled the tSZ effect using two independent techniques.
For smaller overdensities we employ a simplistic biasing model that
rely on a perturbative description of dark matter clustering. The
statistics of the hot gas is than linked to that of the dark matter
using a linear (redshift dependent) biasing scheme. While this type
of modelling is adequate for small overdensities a more elaborate
analytical scheme is required for a detailed description of baryonic
clustering at smaller scales. We consider a halo model based
approach for the gas in collapsed virialised halos The specific
number density and radial profile of these halos are modeled using
Press-Sechter formalism or its variants. These two pictures of
baryonic clustering is complementary to each other. The tSZ angular
power spectrum corresponds to the projected power spectrum of the
baryonic pressure fluctuation and the tSZ angular bispectrum
correspond to the bispectrum of pressure fluctuations projected onto
the surface of the sky.

Few comments on the validity of the perturbative results are in
order. The tSZ power spectrum by and large depends on the one halo
term in the halo modelling. However in the perturbative regime we
are mostly probing the Jean-Scale smoothed gas that is not in
collapsed objects and traces the smoothed large scale dark matter
distribution. To probe the large scale SZ effect removing X-ray
bright clusters can reduce contributions from collapsed halos. Thus
the effect captured by the linear biasing scheme should be
understood as a signal in blank fields where such clusters are
absent. It provides the lower limit of SZ effect from large scale
structure distribution. Though the diffuse component of the tSZ
effect is beyond WMAP detection threshold the situation may improve
with future data sets such as Planck (\cite{Han05}, also see \cite{Joudaki10}). 

The tSZ effect is intrinsically non-Gaussian. While the tSZ power
spectrum is sensitive to the amplitude of the density fluctuations
the higher order statistics such as skewness of tSZ effect can be
used to separate the pressure bias from the amplitude of the density
fluctuations. The skewness is related to the bispectrum of the tSZ
effect. The individual modes of a specific bispectrum is defined by
the triplets of the harmonic numbers $(l_1,l_2,l_3)$. However
individual modes of the bispectrum has low signal to noise and may
not be easy to estimate from a noisy data. In a recent study
\cite{MuHe10} advocated using a spectra correspond to the higher
order spectra in general and bispectrum in particular. The skew
spectrum is the lowest such spectrum that is constructed from
bispectrum. The skew-spectra can retain some of the shape dependence
in bispectrum without compressing it to a single number. The three
different skew-spectra that we introduced in this paper are
generalizations of the ordinary skew-spectrum that was originally
introduced in \cite{MuHe10} and later used in the context of weak
lensing and CMB for morphological analysis. These three different
skew-spectra can also be used to study the morphology of the tSZ
maps as described by the MFs. They associate varying weights to
individual bispectrum modes when constructing the skew-spectra and
carry independent information.

Using near all-sky setup and noise that reflects ongoing CMB
observations such as Planck we study how well the three different
skew-spectra can be estimated from the data. We find that the data
will allow a {\em robust} determination of two of the three
skew-spectra that we have considered with a very high level of
signal to noise. This is true for both perturbative results and the
results that are based on halo model for the entire range of
smoothing angular scales that we have studied. We also find the
estimation of $S^{(2)}$ will be dominated by noise. The high
signal-to-noise for the other two power spectra will allow mode by
mode estimation of each skew-spectra. This can help to differentiate
them from other sources of non-Gaussianity.

The method that we pursue here depends on frequency cleaned tSZ
maps. The tSZ effect can also be studied using cross-correlation
techniques that involve external tracers. Such methods typically
employ mixed bispectra. The results however lack the frequency
information and typically confusion noise dominated. The study of
tSZ using bispectrum from frequncy cleaned maps typically provide
additional signal-to-noise due to the frequency information. In the
absence of frequency information the background CMB plays the role
of intrinsic noise that degrades the signal-to-noise ratio. It is
also interesting to note that removal of tSZ from the CMB maps may
actually help in detection of other sub-dominant effects that we
have not studied here, such as the kinetic Sunyaev Zeldovich effect
(kSZ).

Some of these techniques described here will have wider applicability.
The idea of generalized skew-spectra was shown to be useful in the context of
weak lensing surveys \citep{MWSC11a} and probing primordial non-Gaussianity from
CMB maps \citep{MSC10a}. In this study we have concentrated
on only the leading order terms in the construction of the power spectra
associated with the MFs. However the next to leading order terms
can also be taken into account using the same formalism.
The next to leading order terms will involve kurt-spectra
that generalizes the concept of skew-spetra at fourth order. The four quadruplet of
generalized kurt-spectra are related to the trispectra in a way that is similar
to the relationship of skewspectra and the bispectrum we have considered
in this paper. The generalized kurt-specta and the related generalized
kurtosis can also be extracted from the data using the same PCL approach
which we have discussed here. However the correction to generaised skew-spectra
associated with MFs resulting from the trispctrum are
expected to be negligible compared to the leading contribution
from the generalized skew-spectra.
\section{Acknowledgements}
\label{acknow}
DM acknowledges support
from STFC standard grant ST/G002231/1 at School of Physics and
Astronomy at Cardiff University where this work was completed. SJ and JS
acknowledge support from from the US Department of Education through
GAANN fellowships at UCI. It is a pleasure for DM to thank Alan Heavens, 
Patrick Valageas and Ludo van Waerbeke for many useful discussions.
\bibliography{paper.bbl}

\begin{thebibliography}{}
%
\bibitem[\protect\citeauthoryear{Aghanim, Majumdar, Silk}{2008}]{AMS08}
Aghanim N, Majumdar S., Silk J., 2008, Rept.Prog.Phys., 71, 066902
\bibitem[\protect\citeauthoryear{Bernardeau et al}{2002}]{Bernardreview02}
Bernardeau F., Colombi S., Gaztanaga E., Scoccimarro R., 2002, Phys.Rept.,367, 1
\bibitem[\protect\citeauthoryear{Birkinshaw}{1999}]{Bir99}
Birkinshaw M. 1999, Phys.Rep, 310, 98
\bibitem[\protect\citeauthoryear{Bouchet et al.}{1992}]{B92}
Bouchet F.R.,  Juszkiewicz R., Colombi S., Pellat R., 1992, ApJ, 394, L5 
\bibitem[\protect\citeauthoryear{Bouchet \& Gispert}{1999}]{BG99}
Bouchet F.R., Gispert R., 1999, New Astronomy, 4, 443
\bibitem[\protect\citeauthoryear{Cen \& Ostriker}{1999}]{CO99}
Cen R., Ostriker J.P., 1999, ApJ, 514,1
\bibitem[\protect\citeauthoryear{Cao, Liu \& Fang}{2007}]{Cao07}
Cao L., Liu J.,  Fang L.-Z. 2007, ApJ, 661, 641
\bibitem[\protect\citeauthoryear{Cooray \& Hu}{2001}]{CH01}
Cooray A., Hu W. 2001, ApJ, 548, 7
\bibitem[\protect\citeauthoryear{Cooray}{2000}]{AC1}
Cooray A, 2000, PRD, 62, 103506
\bibitem[\protect\citeauthoryear{Cooray}{2001}]{AC2}
Cooray A, 2001, PRD, 64, 063514
\bibitem[\protect\citeauthoryear{Cooray}{2001}]{Cooray01} 
Cooray A., 2001, PRD, 64, 043516
\bibitem[\protect\citeauthoryear{Cooray, Hu \& Tegmark}{2000}]{CHT00}
Cooray A, Hu W, Tegmark M.,2000,ApJ, 540, 1
\bibitem[\protect\citeauthoryear{Cooray \& Seth}{2002}]{CooSeth02}
Cooray A., Seth R., 2002, Phys. Rep. 372, 1
\bibitem[\protect\citeauthoryear{Cooray, Baumann \& Sigurdson}{2005}]{CBS05}
Cooray, A.; Baumann, D.; Sigurdson, K., Background Microwave Radiation and Intracluster Cosmology. 
Edited by F. Melchiorri and Y. Rephaeli. 
Proceedings of the International School of Physics "Enrico Fermi", 
ISBN 1-58603-585-1
\bibitem[\protect\citeauthoryear{Goldberg \& Spergel}{1999a}]{GS99a}
Goldberg D.M., Spergel D.N., 1999, PRD, 59, 103001
\bibitem[\protect\citeauthoryear{Goldberg \& Spergel}{1999b}]{GS99b}
Goldberg D.M., Spergel D.N., 1999, PRD, 59, 103002
\bibitem[\protect\citeauthoryear{Delabroullie, Cardoso \& Patanchon}{2003}]{DCP03}
Delabrouille J., Cardoso J., Patanchon G., 2003, MNRAS, 330, 807
\bibitem[\protect\citeauthoryear{Edmonds}{1968}]{Ed68}
Edmonds, A.R., Angular Momentum in Quantum Mechanics, 2nd ed. 
rev. printing. Princeton, NJ:Princeton University Press, 1968.
\bibitem[\protect\citeauthoryear{Hadwiger}{1959}]{Hadwiger59}
Hadwiger H. 1959, Normale Koper im Euclidschen raum und ihre topologischen and metrischen
Eigenschaften, Math Z., 71, 124
\bibitem[\protect\citeauthoryear{Fry}{1984}]{F84}
Fry J.N., 1984, ApJ, 279, 499
\bibitem[\protect\citeauthoryear{Hallman et al.}{2009}]{Hal09}
Hallman E.J., O'Shea B.W., Smith B.D., Burns J.O., Norman M.L. 2009, ApJ, 698, 1759
\bibitem[\protect\citeauthoryear{Hallman et al.}{2007}]{Hal07}
Hallman E.J., O'Shea B.W., Burns J.O., Norman M.L., Harkness R., Wagner R., 2007,ApJ, 671,27
\bibitem[\protect\citeauthoryear{Hivon et al.}{2002}]{Hiv} 
Hivon E., G{\'o}rski K.~M., Netterfield C.~B., Crill B.~P., Prunet S., 
Hansen F., 2002, ApJ, 567, 2 
\bibitem[\protect\citeauthoryear{Hansen et al.}{2005}]{Han05} 
Hansen F., Branchini E., Mazzotta P., Cabella P., Dolag K., 2005, MNRAS, 361, 753    
\bibitem[\protect\citeauthoryear{Hikage et al.}{2008}]{Hk08}
Hikage C., Coles P., Grossi M., Moscardini L., Dolag K., Branchini L., Matarrese S.
2008, MNRAS,385,1513
\bibitem[\protect\citeauthoryear{Hikage et al.}{2006}]{HKM06}
Hikage C., Komatsu E., Matsubara T., 2006, ApJ., 653, 11
\bibitem[\protect\citeauthoryear{Hikage, Taruya \& Suto}{2003}]{HTS03}
Hikage C., Taruya A., Suto Y., 2003, Publ.Astron.Soc.Jap, 55, 335
\bibitem[\protect\citeauthoryear{Hikage et al.}{2002}]{Hk02}
Hikage C., et al. Publ.Astron.Soc.Jap. 54 (2002) 707
\bibitem[\protect\citeauthoryear{Hikage et al.}{2003}]{Hk03}
Hikage C., et al. Publ.Astron.Soc.Jap. 55 (2003) 911
\bibitem[\protect\citeauthoryear{Hui}{1999}]{Hui99}
Hui L., 1999, ApJ, 519, L9
\bibitem[\protect\citeauthoryear{Joudaki et al.}{2010}]{Joudaki10}
Joudaki S., Smidt J., Amblard A., Cooray A., 2010, JCAP, 1008, 027
\bibitem[\protect\citeauthoryear{Komatsu \& Seljak}{2001}]{KS02}
Komatsu E.,  Seljak U. 2002, MNRAS, 336,1256
\bibitem[\protect\citeauthoryear{Leach}{2008}]{Leach08}
Leach S.M., et al., A\&A, 491, 597, 2008
\bibitem[\protect\citeauthoryear{Limber}{1954}]{Limb54}
Limber D.N., 1954, ApJ, 119, 665
\bibitem[\protect\citeauthoryear{Lin et al.}{2004}]{Lin04}
Lin K.-Y, Woo T.-P., Tseng Y.-H., Lin L., Chiueh T. 2004, ApJ, 608, L1
\bibitem[\protect\citeauthoryear{LoVerde \& Afshordi}{2008}]{LoAf08}
LoVerde M., Afshordi N. 2008, Phys.Rev.D78, 123506
\bibitem[\protect\citeauthoryear{Mo \& White.}{1996}]{MoWh96}
Mo H.J., Jing Y.P., White S.D.M.  MNRAS, 1996, 284, 189
\bibitem[\protect\citeauthoryear{Mo, Jing \& White}{1997}]{MJW97}
Mo H.J., White S.D.M.  MNRAS, 1997, 282, 347
\bibitem[\protect\citeauthoryear{Munshi \& Heavens}{2010}]{MuHe10}
Munshi D., Heavens A. 2010, MNRAS, 401, 2406
\bibitem[\protect\citeauthoryear{Munshi et al.}{2009}]{Mu09}
Munshi D., Heavens A. Cooray A., Valageas P., 2009, MNRAS (in press).
\bibitem[\protect\citeauthoryear{Munshi, Smidt \& Cooray}{2010}]{MSC10a}
Munshi D. Smidt J. Cooray A. 2010, arXiv:1011.5224 
\bibitem[\protect\citeauthoryear{Munshi et al.}{2011}]{MWSC11a}
Munshi D. van Waerbeke L., Smidt J., Coles P., 2011, arXiv:1103.1876 
\bibitem[\protect\citeauthoryear{Navarro, Frenk \& White}{1996}]{NFW96}
Navarro J., Frenk C., White S.D.M 1996, ApJ, 462, 563
\bibitem[\protect\citeauthoryear{Peebles}{1971}]{Peebles71}
Peebles P.J.E., 1971, Physical cosmology, Princeton Series in Physics.
\bibitem[\protect\citeauthoryear{Persi et al.}{1995}]{Persi95}
Persi F., Spergel D. cen R., Ostriker J., 1995, ApJ, 442,1
\bibitem[\protect\citeauthoryear{Press \& Schechter}{1974}]{PS74}
Press W. \& Sechter P. 1974, Astrophys.J, 187, 425
\bibitem[\protect\citeauthoryear{Rephaeli}{1995}]{Rep95}
Rephaeli Y. 1995, ARA\&A, 33, 541
\bibitem[\protect\citeauthoryear{Refregier et al.}{2000}]{Ref00}
Refregier A., Komatsu E., Spergel D.N., Pen U. -L., 2000, PRD, 61, 123001
\bibitem[\protect\citeauthoryear{Runyan et al.}{2003}]{Run03}
Runyan M.C. et al. 2003 ApJS 149 265
\bibitem[\protect\citeauthoryear{Sunyaev \& Zeldovich}{1980}]{SZ80}
Sunyaev R.A., Zel\'dovich Ya B., 1980, ARA\&A, 18, 537
\bibitem[\protect\citeauthoryear{Sunyaev \& Zeldovich}{1972}]{SZ72}
Sunyaev R.A., Zel\'dovich Ya B., Comments Astrophys. Space Phys, 4, 173
\bibitem[\protect\citeauthoryear{Scoccimarro \& Frieman}{1999}]{SF99}
Scoccimarro, R.; Frieman, J. A., 1999, ApJ, 520, 35S
\bibitem[\protect\citeauthoryear{Scoccimarro \& Couchman}{2001}]{SC01}
Scoccimarro R., Couchman H.M.P., 2001,MNRAS, 325, 1312
\bibitem[\protect\citeauthoryear{da Silva et al.}{1999}]{dasilva99}
da Silva et al., Barbosa A.C., Liddle A.R.,  Thomas P.A., 1999, MNRAS
\bibitem[\protect\citeauthoryear{Refregier \& Teyssier}{2002}]{Ref99}
Refregier A. \& Teyssier, 2002, PhRvD, 66, 043002
\bibitem[\protect\citeauthoryear{Roncarelli et al.}{2007}]{Ron07} 
Roncarelli M., Moscardini L., Borgano S., Dolag K., 2007, MNRAS, 378, 1259
\bibitem[\protect\citeauthoryear{The Planck Collaboration}{2006}]{PC06}
The Planck Collaboration, 2006, astro-ph/0604069
\bibitem[\protect\citeauthoryear{Seljak}{2000}]{Sj00}
Seljak U., MNRAS,2000, 318, 203
\bibitem[\protect\citeauthoryear{Seljak et al}{2001}]{Sel01}
Seljak U., Burwell J., Pen U -L, 2001, PRD, 64, 063001
\bibitem[\protect\citeauthoryear{Smidt et al.}{2010}]{Smidt10}
Smidt J., Joudaki S., Serra P., Amblard A., Cooray A., 2010, PRD, 81, 123528
\bibitem[\protect\citeauthoryear{Springel et al}{2001}]{Spr01}
Springel V., White M. \& Hernquist L., 2001, ApJ, 549,681
\bibitem[\protect\citeauthoryear{White, Hernquist \& Springel}{2002}]{White02}
White M., Hernquist V. \& Springel V., 2002, ApJ, 579,16
\bibitem[\protect\citeauthoryear{Zhang \& Pen}{2001}]{Zp01}
Zhang P. \& Pen U.-L, 2001, ApJ,549,18
\bibitem[\protect\citeauthoryear{Zhang et al}{2004}]{Z04}
Zhang P., Pen U.-L, Trac H., 2004, MNRAS, 355, 451
\bibitem[\protect\citeauthoryear{Zhang \& Seth}{2007}]{ZS07}
Zhang P. \& Seth R.K., 2007, ApJ, 579,16
%
\end{thebibliography}
\end{document}